\documentclass[12pt,onecolumn,journal,draftcls]{IEEEtran}%
\usepackage{amsfonts}
\usepackage{amsmath}
\usepackage{amssymb}
\usepackage{graphicx}%
\newtheorem{theorem}{Theorem}

\newtheorem{lemma}[theorem]{Lemma}

\begin{document}

\title{Relay vs. User Cooperation in Time-Duplexed Multiaccess Networks}
\author{Lalitha Sankar,~\IEEEmembership{Member,~IEEE,} Gerhard
Kramer,~\IEEEmembership{Member,~IEEE,} and~Narayan B.
Mandayam,~\IEEEmembership{Senior Member,~IEEE}\thanks{The work of L.~Sankar
and N.~B.~Mandayam was supported in part by the National\ Science Foundation
under Grant~No.~{\scriptsize ITR-0205362}. The work of G.~Kramer was partially
supported by the Board of Trustees of the University of Illinois Subaward
No.~04-217 under NSF Grant~No.~{\scriptsize CCR-0325673 }and by the Army
Research Office under the ARO grant W911NF-06-1-0182. The material in this
correspondence was presented in part at the IEEE International Symposium on
Information Theory, Adelaide, Australia, Sep. 2005; and at the $45^{th}$
Annual Allerton Conference on Communications, Control, and Computing,
Monticello, IL, Sep. 2007. L. Sankar is with Princeton University. G. Kramer
is with Bell Labs, Alcatel-Lucent. N.~B.~Mandayam is with the WINLAB, Rutgers
University. }}
\pubid{~}
\maketitle

\begin{abstract}
The performance of user-cooperation in a multi-access network is compared to
that of using a wireless relay. Using the total transmit and processing power
consumed at all nodes as a cost metric, the outage probabilities achieved by
dynamic decode-and-forward (DDF) and amplify-and-forward (AF) are compared for
the two networks. A geometry-inclusive high signal-to-noise ratio (SNR) outage
analysis in conjunction with area-averaged numerical simulations shows that
user and relay cooperation achieve a maximum diversity of $K$ and $2$
respectively for a $K$-user multiaccess network under both DDF and AF.
However, when accounting for energy costs of processing and communication,
relay cooperation can be more energy efficient than user cooperation, i.e.,
relay cooperation achieves \textit{coding} (SNR) \textit{gains}, particularly
in the low SNR regime, that override the diversity advantage of user cooperaton.

\end{abstract}

\section{\label{Sec_I}Introduction}

Cooperation results when nodes in a network share their power and bandwidth
resources to mutually enhance their transmissions and receptions. Cooperation
can be induced in several ways. We compare two approaches to inducing
cooperation in a multiaccess channel (MAC) comprised of $K$ sources and one
destination. First, we allow source nodes to forward data for each other and
second, we introduce a wireless relay node when cooperation between the
sources nodes is either undesirable or not possible. We refer to networks
employing the former approach as \textit{user cooperative (UC) networks} and
those employing the latter as \textit{relay cooperative (RC) networks}.

There are important differences between user cooperative and relay networks
that are not easy to analyze from an information-theoretic point of view. For
example, in cooperative networks one likely needs economic incentives to
induce cooperation. On the other hand, relay networks incur infrastructure
costs (see \cite{cap_theorems:SKM04}). While incentives and infrastructure
costs are important issues, we use the total transmit and processing power
consumed for both cooperative and non-cooperative transmissions in each
network as a cost metric for our comparisons. To this end, we model the
processing power as a function of the transmission rate, and thereby the
transmit signal-to-noise ratio (SNR). We also introduce \textit{processing
scale factors} to characterize the ratio of the energy costs of processing
relative to that for transmission. While the processing (energy and chip
density) costs involved in encoding and decoding are complex functions of the
specific communication and computing technologies used, the parametrization we
introduce through scale factors allows us to study the impact of such
processing costs.

We motivate our analysis with examples of wireless devices serving three
different applications \cite[p. 141]{cap_theorems:KramMarYates01}. Consider a
Motorola RAZR GSM mobile phone with a maximum transmit power constraint of 1
(2) W in the 900 (1900) MHz band. With a 3.7 V battery rated at 740 mAh this
device has a capacity of almost 10 kJ of energy resulting in an average talk
time of 4 hours. On the other hand, an Atheros whitepaper
\cite{cap_theorems:Atheros_WP} found that typical 802.11 wireless local area
network (WLAN) interfaces consume 2 to 8 W for active communications.
Furthermore, the transmit power for this device in the range of 20 to 100 mW
is only a small fraction of the processing costs. Finally, consider low-power
sensor devices such as the Berkeley motes. The authors in
\cite{cap_theorems:HienChandBala} model the energy cost per bit for a reliable
1 Mbps link over a distance $d$ and path-loss exponent $\alpha$ by a
transmitter cost of $\mathcal{E}_{tx}=\mathcal{E}_{t}+\mathcal{E}%
_{pa}d^{\alpha}$ where $\mathcal{E}_{t}=0.36$ J/MB is the energy dissipated in
the transmitter electronics and $\mathcal{E}_{pa}=8\times10^{-5}$ J/m$^{2}$/MB
scales the required transmit energy per bit. Accounting for the signal
processing costs at the receiver as $\mathcal{E}_{rx}=1.08$ J/MB, they show
that for distances less than the transition distance of $d=\sqrt
{\mathcal{E}_{t}/\mathcal{E}_{pa}}=67$ m, processing energy cost dominates
transmission cost and vice-versa. In general, the ratio of processing to
transmission power depends on both the device functionality (long distance vs.
local links) and the application (high vs. low rate) supported. Thus,
accounting for both the transmit and processing power (energy) costs in our
comparisons allows us to identify the processing factor\textit{ }regimes where
cooperation is energy efficient.

We consider single-antenna half-duplex nodes and constrain all transmitting
nodes in both networks to time-duplex their transmissions. Thus, in the relay
network each source cooperates with the relay over two-hops where in the first
hop the source transmits while the relay listens and in the second hop both
the source and relay transmit. For the user cooperative network, for $K>2$ we
consider the cooperative schemes of \textit{two-hop}, where the cooperating
users of any source transmit in the second hop, and \textit{multi-hop}, where
the cooperating users transmit sequentially in time. We assume that
transmitters do not have channel state information (CSI) and compare the
outage performance of the two networks as a function of the transmit SNR at
each user for the cooperative strategies of \textit{dynamic }%
decode-and-forward (DDF) \cite{cap_theorems:AGS01} and amplify-and-forward
(AF). We present upper and lower bounds on the outage probability of DDF and
AF for both networks and compare their outage performance via a \textit{coding
}(\textit{SNR})\textit{ gain} \cite{cap_theorems:JNL01}. For single-antenna
nodes, the maximum DDF\ and AF diversity for two-hop relaying is $2$
\cite{cap_theorems:AGS01}. For the two-hop user cooperative network, we show
that, if relay selection is allowed, AF achieves a maximum diversity of $2$.
Further, we also show that, except for a \textit{clustered} geometry where the
maximum diversity approaches $K$, DDF also achieves a maximum diversity of $2$
for this network. On the other hand, when users cooperate using a $K$-hop
scheme, our bounding analysis agrees with the earlier results that both DDF
\cite{cap_theorems:AGS01} and AF
\cite{cap_theorems:JNL01,cap_theorems:LalithaSankar} achieve a maximum
diversity of $K$.

The coding gains achieved are in general a function of the transmission
parameters and network geometry. In an effort to generalize such results, we
present an \textit{area-averaged} numerical comparison. Specifically, we
consider a sector of a circular area with the destination at the center, a
fixed relay position, and the users randomly distributed in the sector. We
remark that this geometry encompasses a variety of centralized network
architectures such as wireless LAN, cellular, and sensor networks. Our
analytical and numerical results demonstrate the effect of processing power in
cooperation and are summarized by the following observations: i) user
cooperation can achieve higher diversity gains than relay cooperation but at
the expense of increased complexity and ii) relay cooperation achieves larger
coding gains when we account for the energy costs of cooperation, thus
diminishing the effect of the diversity gains achieved by user cooperation.

This paper is organized as follows. In Section \ref{Sec_2}, we present the
network and channel models and develop a power-based cost metric. In Section
\ref{Sec_3}, we present the outage approximations for the DDF and AF
strategies for both networks. In Section \ref{Sec_4}, we present the numerical
results. We conclude in Section \ref{Sec_5}.

\section{\label{Sec_2}Channel and Network Models}

\subsection{\label{Sec2_SS1}Network Model}

Our networks consist of $K$ users (source nodes) numbered $1,2,\ldots,K$ and a
destination node $d$. For the relay network there is one additional node, the
relay node $r$. We impose a \textit{half-duplex} constraint on every node,
i.e., each node can be in one of two modes, \textit{listen} (\textit{L}) or
\textit{transmit} (\textit{T}) (\textit{LoT}). We write $\mathcal{K}%
=\{1,2,\ldots,K\}$ for the set of users and $\mathcal{T}=\mathcal{K\cup}\{r\}$
for the set of transmitters in the relay network.

Let $X_{k,i}$ be the transmitted signal (channel input) at node $k$ at time
$i$, $i=1,$ $2,$ $\ldots,n$. We model the wireless multiaccess links under
study as additive Gaussian noise channels with fading. For such channels, the
received signal (channel output) at node $m$ at time $i$ is
\begin{equation}
Y_{m,i}=\left\{
\begin{array}
[c]{ll}%
\left(  \sum\limits_{k\neq m}H_{m,k,i}X_{k,i}\right)  +Z_{m,i} & M_{m,i}=L\\
0 & M_{m,i}=T
\end{array}
\right.  \label{Gen_MAC_defn}%
\end{equation}
where the $Z_{m,i}$ are independent, proper, complex, zero-mean, unit variance
Gaussian noise random variables, $M_{m,i}$ is the half-duplex mode at node
$m$, and $H_{m,k,i}$ is the complex fading gain between transmitter $k$ and
receiver $m$ at time $i$. Note that for both networks as well as the
(non-cooperative) MAC, $M_{d,i}=L$, for all $i$. Further, for the relay
network and the MAC, we also have $M_{k,i}=T$, for all $i$ and for all
$k\in\mathcal{K}$. We assume that over all $n$ uses of the channel, the
transmitted signals in both networks are constrained in power as
\begin{equation}%
\begin{tabular}
[c]{ll}%
$\sum\limits_{i=1}^{n}E\left\vert X_{k,i}\right\vert ^{2}\leq nP_{k}$ &
$k\in\mathcal{T}$.
\end{tabular}
\ \ \ \ \ \ \ \ \ \ \label{MAC_pwr_constrant}%
\end{equation}
Throughout the sequel we assume that all transmitters use independent Gaussian
codebooks with asymptotically large codelengths and the total transmission
bandwidth is unity. Further, we assume that the modes $M_{k,i}$ are known by
all nodes. Finally, we use the usual notation for entropy and mutual
information \cite{cap_theorems:CTbook} and take all logarithms to the base 2
so that our rate units are bits/channel use. We write random variables (e.g.
$H_{k}$) with uppercase letters and their realizations (e.g. $h_{k}$) with the
corresponding lowercase letters and use the notation $C(x)$ $=$ $\log(1+x)$
where the logarithm is to the base 2. Finally, throughout the sequel we use
the words \textquotedblleft user\textquotedblright\ and \textquotedblleft
source\textquotedblright\ interchangeably.

\subsection{\label{Hier_HDMARC_defn}Relay Cooperative Network}

\begin{figure*}[tbp] \centering
{\includegraphics[
height=3.4861in,
width=5.2295in
]%
{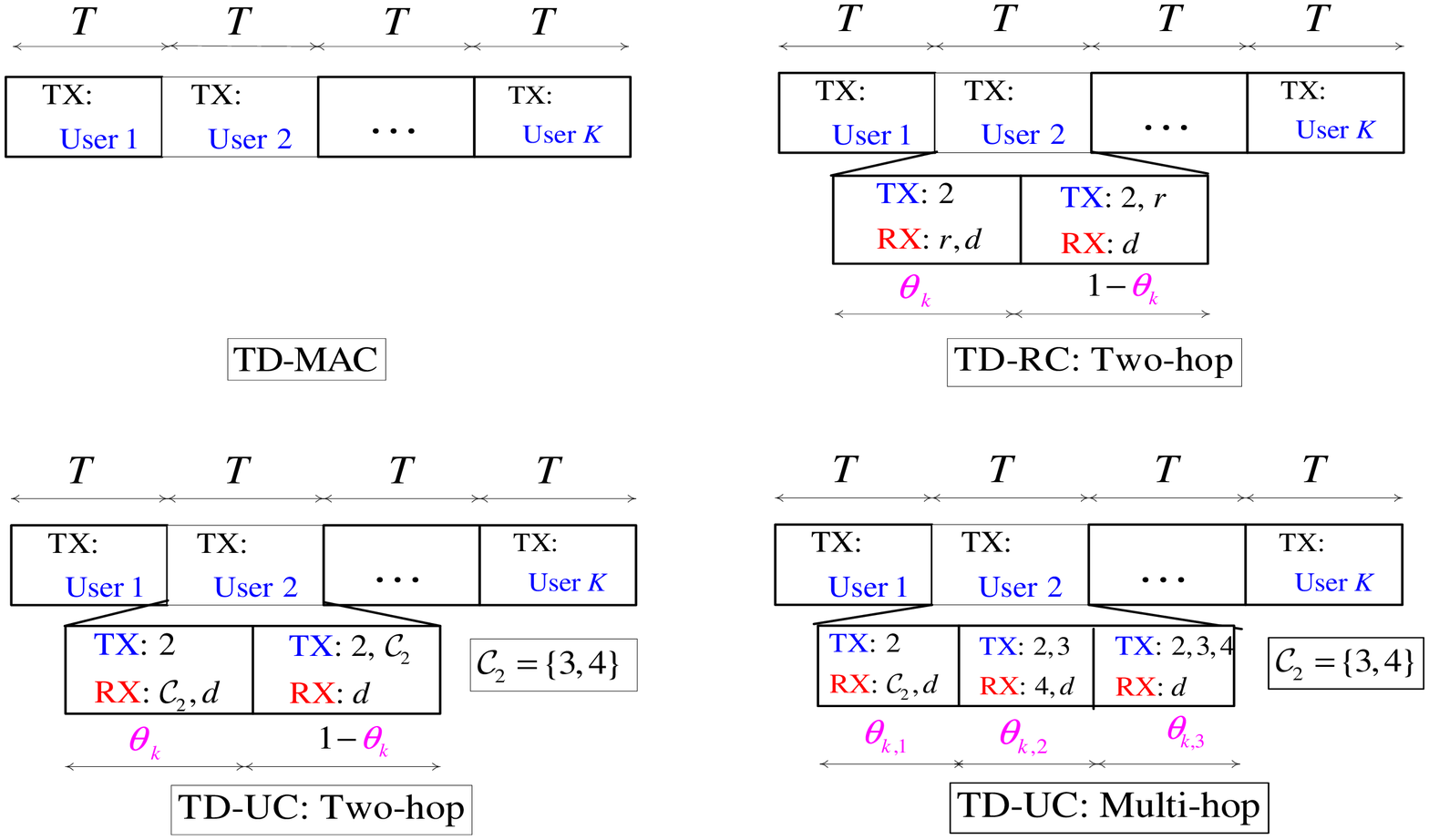}%
}%
\caption{Time-duplexed transmission schemes for the MARC, the MAC-GF, and the MAC.}\label{Fig_Slot_Schemes}%
\end{figure*}

The relay cooperative (RC) network with $K+1$ inputs $X_{k,i}$, $k\in
\mathcal{T}$, and two outputs $Y_{r,i}$ and $Y_{d,i}$ given by
(\ref{Gen_MAC_defn}) is typically modeled as a Gaussian multiaccess relay
channel (MARC) \cite{cap_theorems:KGG_IT,cap_theorems:SKM04}. We consider a
time-duplexed relay cooperative (TD-RC) model where each source transmits over
the channel for a period $T=1/K$ of the total time (see Fig.
\ref{Fig_Slot_Schemes}). Further, the transmission period of source $k$, for
all $k$, is sub-divided into two slots such that the relay listens in first
slot and transmits in the second slot. We denote the time fractions for the
two slots as $\theta_{k}$ and $\overline{\theta}_{k}=1-\theta_{k}$ for user
$k$ such that $\theta_{k}=\Pr\left(  M_{r}=L\right)  =1-\Pr\left(
M_{r}=T\right)  $ where the duration, $\theta_{k}$, of the relay mode $M_{r}$
can be different for different $k$. The time-duplexed two-hop scheme for the
RC nework is illustrated in Fig. \ref{Fig_Slot_Schemes} for user $2$. Also
shown is the slot structure for a time-duplexed MAC\ (TD-MAC). Time-duplexing
thus simplifies the analysis for each user to that for a single-source relay
channel in each period $T$. We assume that the relay uses negligible resources
to communicate its mode transition to the destination. We also assume that, to
minimize outage, the transmitters use all available power for transmission
subject to (\ref{MAC_pwr_constrant}). Thus, in the $k^{th}$ time period, for
all $k$, user $k$ and the relay transmit at power $\overline{P}_{k}=KP_{k}$
and $\overline{P}_{r}=P_{r}/\overline{\theta}_{k}$, respectively, where
$\overline{\theta}_{k}=1-\theta_{k}$. Finally, throughout the analysis we
assume that $P_{r}$ is proportional to $P_{k}$.

\subsection{User Cooperative Network}

In a user cooperative (UC) network, there is a combinatorial explosion in the
number of ways one can duplex $K$ sources over their half-duplex states. We
present two transmission schemes that allow each user to be aided by an
arbitrary number of users, up to $K-1$. In both schemes the users time-duplex
their transmissions; the two schemes differ in the manner the period $T$ is
further sub-divided between the transmitting and the cooperating users.

We first consider a \textit{two-hop scheme} such that the period over which
user $k$, for all $k$, transmits is sub-divided into two slots. In the first
slot only user $k$ transmits while in the second slot both user $k$ and the
set $\mathcal{C}_{k}\subseteq\mathcal{K}\backslash\{k\}$ of users that
cooperate with user $k$ transmit. This is shown in Fig. \ref{Fig_Slot_Schemes}
for user $2$ and $\mathcal{C}_{2}=\left\{  3,4\right\}  $. We remark that this
scheme has the same number of hops as the TD-RC network except now user $k$
can be aided by more than one user in $\mathcal{C}_{k}$. We write $\theta_{k}$
and $1-\theta_{k}$ to denote the time fractions associated with the first and
second slots of user $k$ such that $\theta_{k}=\Pr\left(  M_{j}=L\right)
=1-\Pr\left(  M_{j}=T\right)  $ for all $j\in\mathcal{C}_{k}$.

We also consider a \textit{multi-hop} scheme where the total transmission time
for source $k$ is divided into $L_{k}$ slots, $1\leq L_{k}\leq K$, where
$L_{k}=\left\vert \mathcal{C}_{k}\right\vert +1$. Specifically, in each
time-slot, except the first slot where only user $k$ transmits, one additional
user cooperates in the transmission until all $L_{k}$ users transmit in slot
$L_{k}$. When the cooperating users decode their received signals, we assume
that the users are ordered in the sense that the new user that cooperates in
the $l^{th}$ fraction is the first user that can decode the message when the
$l$ cooperating users are transmitting. We denote the $l^{th}$ time fraction
for user $k$ as $\theta_{k,l}$, $l=1,2,\ldots,L$ (see Fig.
\ref{Fig_Slot_Schemes} for user $2$ with $\mathcal{C}_{2}=\left\{
3,4\right\}  $).\ We refer to this model as time-duplexed user cooperation or
simply TD-UC.

User $k$ transmits at power
\begin{equation}
\overline{P}_{k}=P_{k}\cdot K\left/  \left(  N_{k}+1\right)  \right.
\label{UC_Khop_Pwr}%
\end{equation}
where $N_{k}\leq K-1$ is the total number of users whose messages are
forwarded by user $k$. Further, for the two-hop scheme, in those sub-slots
where user $k$ acts as a cooperating node, its transmission power is scaled by
the appropriate $\overline{\theta}_{k}$. The energy consumed in every
cooperative slot is therefore exactly given by (\ref{UC_Khop_Pwr}). Let
$\pi_{k}\left(  \cdot\right)  $ be a permutation on $\mathcal{C}_{k}$ such
that user $\pi_{k}\left(  l\right)  $ begins its transmissions in the fraction
$\theta_{k,l}$, for all $l=2,3,\ldots,L_{k}$, and $\pi_{k}\left(  1\right)
=k$. Thus, when user $k$ acts as a cooperating node for user $j$, $j\not =k$,
such that $\pi_{j}(l)=k$ for some $l>1$, its power $\overline{P}_{k}$ in
(\ref{UC_Khop_Pwr}) is scaled by the total fraction for which it transmits for
user $j$, i.e., $%
{\textstyle\sum\nolimits_{m=l}^{L_{j}}}
\theta_{j,m}$. We assume that a cooperating node or relay uses negligible
resources to communicate its transition from one mode to another to the
destination as well as other cooperating nodes. For AF we assume equal length
slots and consider symbol-based two-hop and multi-hop schemes.

Finally, throughout the sequel, we assume that due to lack of CSI at the
transmitters, the transmitters do not vary power as a function of channel
states. Furthermore, each user uses independent Gaussian in each transmitting
fraction. Thus, for e.g., for two-hop TD-RC and TD-UC networks under AF, user
$k$ transmits independent codebooks with the same power in the two fractions.
Similarly, for the multi-hop TD-UC network under AF, subject to
(\ref{UC_Khop_Pwr}), user $k$ transmits independent signals with the same
power in all $L_{k}$ fractions.

\subsection{Cost Metric: Total Power}

We use the total power consumed by all the nodes as a cost metric for
comparisons. Observe that in addition to its transmit power a node also
consumes processing power, i.e., in encoding and decoding its transmissions
and receptions, respectively. Further, in addition to its own transmission and
processing costs, a node that relays consumes additional power in encoding and
decoding packets for other nodes. We model these costs by defining encoding
and decoding variables $\eta_{k}$ and $\delta_{k}$, respectively, and write
the power required to process the transmissions of node $j$ at node $k$ as
\begin{equation}%
\begin{array}
[c]{cc}%
P_{k,j}^{proc}=P_{k,0}^{proc}+\left(  \eta_{k}I_{k}^{enc}\left(  j\right)
+\delta_{k}I_{k}^{dec}\left(  j\right)  \right)  \cdot f\left(  R_{j}\right)
& \text{for all }k\in\mathcal{T},j\in\mathcal{K}%
\end{array}
\end{equation}
where $P_{k,j}^{proc}$ is the power required by user $k$ to cooperate with
user $j$, $I_{k}^{enc}\left(  j\right)  $ and $I_{k}^{dec}\left(  j\right)  $
are indicator functions that are set to $1$ if user $k$ encodes and decodes,
respectively, for user $j$, $P_{k,0}^{proc}$ is the minimum processing power
at user $k$ which is in general device and protocol dependent, and $f(R_{j})$
is a function of the transmission rate $R_{j}$ in bits/sec at user $j$. The
unitless variables $\eta_{k}$ and $\delta_{k}$ quantify the ratio of
processing to transmission power at user $k$ to encode and decode a bit,
respectively. For example, a relay node that uses DDF consumes power for
overhead, encoding, and decoding costs while a relay node using AF only has
overhead costs. Note that for the relay node, we have $P_{r,r}^{proc}%
=P_{r,0}^{proc}$ which accounts for the costs of simply operating the relay.
Thus, for the examples in Section \ref{Sec_I}, we have $\eta<1$ and $\delta<1$
for the RAZR phone, $\eta>>1$ and $\delta>>1$ for the Atheros LAN card, and
$\eta$ and $\delta$ determined by the cross-over distance for the Berkeley
motes. In general, the processing cost function $f$ depends on the encoding
and decoding schemes used as well as the device functionality. For simplicity,
we choose $f$ as
\begin{equation}%
\begin{array}
[c]{cc}%
f\left(  R_{k}\right)  =R_{k} & \text{for all }k.
\end{array}
\label{Comp_f_choice}%
\end{equation}
Finally, we assume that the destination in typical multiaccess networks such
as cellular or many-to-one sensor networks has access to an unlimited energy
source and ignore its processing costs. We write the total power consumed on
average (over all channel uses) at node $k$, $k\in\mathcal{T}$, as
\begin{equation}
P_{k,tot}=\left\{
\begin{array}
[c]{ll}%
P_{k}+P_{k,k}^{proc}+\sum\limits_{j\in\mathcal{K},j\not =k}I_{k}%
(j)P_{k,j}^{proc} & k\in\mathcal{K}\\
P_{k}+\sum\limits_{j\in\mathcal{K}}I_{k}(j)P_{k,j}^{proc} & k=r
\end{array}
\right.  \label{Tot_pwr_costs}%
\end{equation}
where $I_{k}(j)$ is an indicator function that takes the value $1$ if node $k$
cooperates with node $j$. For user $k,$ the first $P_{k,k}^{proc}$ term in
(\ref{Tot_pwr_costs}) corresponds to the power used to process its own message
while the second summation term accounts for the power node $k$ incurs in
cooperating with all other source nodes. Note that at high SNR, i.e., high
$P_{k}$ for all $k$, the dominating term in (\ref{Tot_pwr_costs}) is $P_{k}$
since $P_{k,0}^{proc}$ is usually a constant and $R_{j}$ increases
logarithmically in $P_{j}$, for all $k,j\in\mathcal{K}$. The total power
consumed by all transmitting nodes in each network is given as
\begin{equation}
P_{tot}=\left\{
\begin{array}
[c]{ll}%
\sum\limits_{k\in\mathcal{K}}P_{k,tot} & \text{TD-MAC\ or UC}\\
\sum\limits_{k\in\mathcal{T}}P_{k,tot} & \text{RC.}%
\end{array}
\right.  \label{Ptot_all_nets}%
\end{equation}

\subsection{\label{Fading_Mods}Fading Models}

We model the fading gains as $H_{m,k,i}=A_{m,k,i}\left/  d_{m,k}^{\gamma
/2}\right.  $ where $d_{m,k}$ is the distance between the $m^{th}$ receiver
and the $k^{th}$ source, $\gamma$ is the path-loss exponent, and the
$A_{m,k,i}$ are jointly independent identically distributed (i.i.d.)
zero-mean, unit variance proper, complex Gaussian random variables. We assume
that the fading gain $H_{m,k,i}$ is known only at receiver $m$. We also assume
that $H_{k,m,i}$ remains constant over a coherence interval and changes
independently from one coherence interval to another. Further, the coherence
interval is assumed large enough to transmit a codeword from any transmitter
and all its cooperating nodes or relay. Finally, we also assume that the
fading gains are independent of each other and independent of the transmitted
signals $X_{k,i}$, for all $k\in\mathcal{T}$ and $i$.

\section{\label{Sec_3}Geometry-inclusive Outage Analysis}

We compare the outage performance of the user and relay cooperative networks
via a limiting analysis in SNR of the outage probabilities achieved by DDF and
AF. Such an analysis enables the characterization of two key parameters,
namely, the diversity order and the coding gains, which correspond to the
slope and the SNR intercept, respectively, of the log-outage vs. SNR\ in dB
curve \cite{cap_theorems:JNL01}. In \cite{cap_theorems:JNL01}, Laneman
develops bounds on the DF and AF outage probabilities for a relay channel
where the source and the relay transmit on orthogonal channels. In
\cite{cap_theorems:AGS01}, the authors introduce a DDF strategy where the
cooperating node/relay remains in the \textit{listen} mode until it
successfully decodes its received signal from the source. The authors show
that, for both two-hop and multi-hop relay channels, DDF achieves the
diversity-multiplexing tradeoff (DMT) performance \cite{cap_theorems:ZT01} of
an equivalent MIMO channel for small multiplexing gains. In an effort to
quantify the diversity and the effect of geometry, we present
geometry-inclusive upper and lower bounds on the DDF and AF outage probability
for TD-RC and two-hop and multi-hop TD-UC networks. We summarize the results
here and develop the detailed analyses in the Appendices.

\subsection{\label{Comp_S3SS1}Dynamic-Decode-and-Forward}

\subsubsection{TD-RC}

In general, obtaining a closed form expression for the outage probability of
each user is not straightforward. Suppose that $P_{r}=\lambda\overline{P}_{k}$
for some constant $\lambda$ and recall that $\overline{P}_{r}=\left.
P_{r}\right/  \overline{\theta}_{k}$. In Appendix \ref{Comp_App2}, we develop
upper and lower bounds on the DDF outage probability $P_{o}^{(k)}$ of user $k$
transmitting at a fixed rate $R_{k}$, for all $k$, as
\begin{equation}
P_{o,2\times1}\leq P_{o}^{(k)}\leq\left[  \frac{(2^{R_{k}/\overline{\theta
}_{k}^{\ast}}-1)^{2}\overline{\theta}_{k}^{\ast}}{(2^{R_{k}}-1)^{2}}%
+\frac{2d_{r,k}^{\gamma}(2^{R_{k}/\theta_{k}^{\ast}}-1)^{2}}{d_{d,r}^{\gamma
}(2^{R_{k}}-1)^{2}}\right]  \cdot\frac{(2^{R}-1)^{2}d_{d,k}^{\gamma}%
d_{d,r}^{\gamma}}{2\lambda\overline{P}_{k}^{2}}+O\left(  \overline{P}_{k}%
^{-3}\right)  \label{MARC_DDF_PoutB}%
\end{equation}
where $P_{o,2\times1}$ is the outage probability of a $2\times1$
\textit{distributed} MIMO channel whose $i^{th}$ transmit antenna is at a
distance $d_{d,i}$, $i=k,r$, from the destination and $\theta_{k}^{\ast}%
\in\left(  0,1\right)  $ is a fraction chosen to upper bound $P_{o}^{\left(
k\right)  }$. The notation $O\left(  x\right)  $ in (\ref{MARC_DDF_PoutB})
means that there is a positive constant $M$ such that the $O\left(  x\right)
$ term is upper bounded by $M\left\vert x\right\vert $ for all $x\geq x_{0}$.
In Appendix \ref{Comp_App2}, we show that
\begin{equation}
P_{o,2\times1}=(2^{R_{k}}-1)^{2}d_{d,k}^{\gamma}d_{d,r}^{\gamma}\left/
2\overline{\lambda P}_{k}^{2}\right.  +O\left(  \overline{P}_{k}^{-3}\right)
. \label{DDF_PMIMO_UBK2}%
\end{equation}
Thus, from (\ref{MARC_DDF_PoutB}) and (\ref{DDF_PMIMO_UBK2}) we see that for a
fixed rate transmission, the maximum diversity achieved by DDF is $2$, as
predicted by the DMT analysis for DDF in \cite[Theorem 4]{cap_theorems:AGS01}.
Comparing (\ref{MARC_DDF_PoutB}) and (\ref{DDF_PMIMO_UBK2}), we further see
that the bracketed expressions on the right side of the inequality in
(\ref{MARC_DDF_PoutB}) upper bounds the coding gains by which $P_{o}^{(k)}$
differs from the MIMO lower bounds.

\subsubsection{TD-UC -- Two-Hop}

The outage analysis for the two-hop TD-RC network can be extended to the
two-hop TD-UC network. In Appendix \ref{Comp_App2}, for sufficiently large
power $P_{k}$, we bound $P_{o}^{(k)}$ as (see \ref{A3_Po_MIMO}) and
(\ref{A3Pub}))%
\begin{equation}
P_{o,L_{k}\times1}\leq P_{o}^{(k)}\leq K_{2}\cdot\frac{\left(  2^{R_{k}%
}-1\right)  ^{L_{k}}\prod_{j\in\mathcal{S}_{k}}d_{d,j}^{\gamma}}{\left(
L_{k}!\right)  \left(  \overline{P}_{k}\right)  ^{L_{k}}\prod\limits_{j\in
\mathcal{S}_{k}}\lambda_{j}}+O\left(  \overline{P}_{k}^{-L_{k}-1}\right)
\label{GF_Po_final}%
\end{equation}
where $\lambda_{j}=\overline{P}_{j}/\overline{P}_{k}$ for all $j\in
\mathcal{S}_{k}=\mathcal{C}_{k}\cup\left\{  k\right\}  $, $\theta_{k}^{\ast
}\in\left(  0,1\right)  $, $P_{o,L_{k}\times1}$ is the outage probability of a
$L_{k}\times1$ \textit{distributed} MIMO channel whose $i^{th}$ transmit
antenna is at a distance $d_{d,i}$, $i=1,2,\ldots,L_{k}$, from the destination
such that
\begin{equation}
P_{o,L_{k}\times1}=\frac{\left(  2^{R_{k}}-1\right)  ^{L_{k}}}{\left(
L_{k}!\right)  \left(  \overline{P}_{k}\right)  ^{L_{k}}}\prod_{j\in
\mathcal{S}_{k}}\frac{d_{d,j}^{\gamma}}{\lambda_{j}}+O\left(  \overline{P}%
_{k}^{-L_{k}-1}\right)  \label{PoMIMO_Lk1}%
\end{equation}
and
\begin{equation}
K_{2}=\left[  \frac{\left(  2^{R_{k}/\overline{\theta}_{k}^{\ast}}-1\right)
^{L_{k}}\left(  \overline{\theta}_{k}^{\ast}\right)  ^{L_{k}-1}}{\left(
2^{R_{k}}-1\right)  ^{L_{k}}}+\frac{(2^{R_{k}/\theta_{k}^{\ast}}-1)^{2}\left(
\sum_{j\in C_{k}}d_{j,k}^{\gamma}\right)  \left(  L_{k}!\right)  \left(
\overline{P}_{k}\right)  ^{L_{k}-2}}{\left(  2^{R_{k}}-1\right)  ^{L_{k}%
}\left(  \prod_{j\in\mathcal{C}_{k}}d_{d,j}^{\gamma}/\lambda_{j}\right)
}\right]  . \label{UC2hop_K2}%
\end{equation}
Note that for $L_{k}=2$, our analysis simplifies to the outage analysis for
the TD-RC network. For $L_{k}>2$, comparing the two terms in the right-hand
sum in (\ref{UC2hop_K2}), we see that a lower bound on the diversity from the
first and second terms are $L_{k}$ and $2$, respectively. In fact, the first
term dominates only when
\begin{equation}
\left(  \sum_{j\in C_{k}}d_{j,k}^{\gamma}\right)  \leq\frac{\left(
2^{R_{k}/\overline{\theta}_{k}^{\ast}}-1\right)  ^{L_{k}-2}}{\left(
L_{k}!\right)  \left(  \overline{P}_{k}\right)  ^{L_{k}-2}}\cdot\frac{\left(
\prod_{j\in\mathcal{S}_{k}}d_{d,j}^{\gamma}/\lambda_{j}\right)  }%
{d_{d,k}^{\gamma}}. \label{MACGF_2h_distCon}%
\end{equation}
Thus, for a given $P_{k}$, for all $k$, achieving the maximum diversity
$L_{k}$ requires that user $k$ and its cooperating users in $\mathcal{C}_{k}$
are clustered close enough to satisfy (\ref{MACGF_2h_distCon}). Thus, the
maximum DDF diversity for a two-hop cooperative network does not exceed that
of TD-RC except when user $k$ and its cooperating users are \textit{clustered}%
, i.e., the inter-node distances satisfy (\ref{MACGF_2h_distCon}). We
illustrate this distance-dependent behavior in Section \ref{Sec_4}.

\subsubsection{TD-UC -- Multi-Hop}

Recall that $\pi_{k}\left(  \cdot\right)  $ is a permutation on $\mathcal{C}%
_{k}$ such that user $\pi_{k}\left(  l\right)  $ begins its transmissions in
the fraction $\Theta_{k,l}$, for all $l=2,3,\ldots,L_{k}$, and $\pi_{k}\left(
1\right)  =k$. Unlike the two-hop case where $\Theta_{k}$ is dictated by the
node with the worst receive SNR, the fraction $\Theta_{k,l}$, for
$l=1,2,\ldots,L_{k}-1$, is the smallest fraction that ensures that at least
one cooperating node, denoted as $\pi_{k}\left(  l+1\right)  $, decodes the
message from user $k$. In general, developing closed form expressions for
$P_{o}^{(k)}$ is not straightforward. In Appendix \ref{Comp_App4}, we lower
bound $P_{o}^{\left(  k\right)  }$ by the MIMO\ outage probability,
$P_{o,L_{k}\times1}$ and use the CDF of $\Theta_{k,l}$, for all $l$, to upper
bound $P_{o}^{\left(  k\right)  }$ for any $0<\theta_{k,l}^{\ast}<1$, for all
$l$, as (see (\ref{A4PubSum}))%
\begin{equation}
P_{o}^{\left(  k\right)  }\leq\frac{\left(  2^{R_{k}}-1\right)  ^{L_{k}}%
}{\left(  L_{k}!\right)  \left(  \overline{P}_{k}\right)  ^{L_{k}}}\left(
\prod_{j=1}^{L_{k}}\frac{d_{d,\pi_{k}\left(  j\right)  }^{\gamma}}%
{\lambda_{\pi_{k}\left(  j\right)  }}\right)  \cdot\left[  K_{c}+K_{d}\right]
+O\left(  \overline{P}_{k}^{-L_{k}-1}\right)
\end{equation}
where the constants $K_{c}$ and $K_{d}$ are given by (\ref{A4_KcKd}) in
Appendix \ref{Comp_App4}. Our analysis shows that DDF achieves a maximum
diversity of $L_{k}$ for a $L_{k}$-hop TD-UC network.

\subsection{\label{Comp_S3SS2}Amplify-and-Forward}

A cooperating node or a relay can amplify its received signal and forward it
to the destination; the resulting AF strategy is appropriate for nodes with
limited processing capabilities. We present the outage bounds for the two-hop
TD-RC and TD-UC and the $L_{k}$-hop TD-UC networks. We assume $\theta_{k}=1/2$
and $\theta_{k,l}=1/L_{k}$, $l=1,2,\ldots,L_{k}$, for the two-hop and $L_{k}%
$-hop schemes, respectively.

\subsubsection{TD-RC and TD-UC -- Two-hop}

We first consider a two-hop AF protocol where only user $k$ transmits in the
first fraction and both user $k$ and its cooperating users (TD-UC) or relay
(TD-RC) transmit in the second fraction. User $k$ transmits with a different
codebook in the first and second fractions. The outage analysis for the
two-hop TD-RC network, i.e., $\left\vert \mathcal{C}_{k}\right\vert =1$, is
the same as that developed for the half-duplex relay channel in
\cite{cap_theorems:NBK01}. Recall that due to lack of transmit CSI, we assume
no power control and independent Gaussian codebooks in each transmit fraction
at user $k,$ for all $k$. For the TD-UC network, i.e., $L_{k}\geq2$, where all
$L_{k}-1$ cooperating nodes amplify and forward their received signals in the
second fraction, the received and transmitted signals $\left(  Y_{d,1}%
,Y_{d,2}\right)  $ and $\left(  X_{k,1},X_{k,2}\right)  $, respectively, in
the two fractions are
\begin{equation}%
\begin{bmatrix}
Y_{d,1}\\
Y_{d,2}%
\end{bmatrix}
=%
\begin{bmatrix}
H_{d,k} & 0\\
\sum\limits_{j\in\mathcal{C}_{k}}c_{j}H_{d,j}H_{j,k} & H_{d,k}%
\end{bmatrix}%
\begin{bmatrix}
X_{k,1}\\
X_{k,2}%
\end{bmatrix}
+%
\begin{bmatrix}
Z_{d,1}\\
Z_{d,2}^{\prime}%
\end{bmatrix}
\label{AF_2hop_Yd}%
\end{equation}
where
\begin{align}
\left\vert c_{j}\right\vert  &  =\left(  2\overline{P}_{j}/\left/  \left\vert
H_{j,k}\right\vert ^{2}\overline{P}_{k}+1\right.  \right)  ^{1/2}%
\label{AF_cj}\\
Z_{d,2}^{\prime}  &  =\left(
{\textstyle\sum\nolimits_{j\in\mathcal{C}_{k}}}
c_{j}H_{d,j}Z_{j,k}\right)  +Z_{d,2}\label{AF_Zprime}\\
c_{s}^{2}  &  =1+\sum\nolimits_{j\in\mathcal{C}_{k}}\left\vert c_{j}%
H_{d,j}\right\vert ^{2}. \label{AF_cs}%
\end{align}
and $Z_{j,k}$, $Z_{d,1}$, and $Z_{d,2}$ are i.i.d. Gaussian noise variables.
Scaling $Y_{d,2}$ by $c_{s}$ to set $E\left[  \left\vert Z_{d,2}^{\prime
}\right\vert ^{2}\right]  =1$, the outage $P_{o}^{(k)}$ is given as
\begin{equation}
P_{o}^{(k)}=\Pr\left(  \frac{1}{2}C\left(  \left\vert H_{d,k}\right\vert
^{2}\overline{P}_{k}\left(  1+\frac{1}{c_{s}}\right)  +\frac{\overline{P}_{k}%
}{c_{s}^{2}}\left\vert \sum\limits_{j\in\mathcal{C}_{k}}\frac{c_{j}}{c_{s}%
}H_{d,j}H_{j,k}\right\vert ^{2}\right)  <R_{k}\right)  \label{AF_2hop_Pout}%
\end{equation}
where the pre-log factor of $1/2$ is a result of $\theta_{k}=1/2$. For
$\left\vert \mathcal{C}_{k}\right\vert >1$, the terms in (\ref{AF_2hop_Pout})
with cross products $H_{d,j}H_{j,k}$ may not add constructively. Accordingly,
we lower bound $P_{o}^{(k)}$ by the outage probability of a $L_{k}\times1$
MIMO channel where all but one of the antennas transmit the same signal, i.e.,%
\begin{equation}
P_{o}^{(k)}\geq\Pr\left(  C\left(  \left\vert H_{d,k}\right\vert ^{2}%
\overline{P}_{k}+\overline{P}_{k}\left\vert
{\textstyle\sum\nolimits_{j\in\mathcal{C}_{k}}}
H_{d,j}\right\vert ^{2}\right)  <R_{k}\right)  =\frac{\left(  2^{R_{k}%
}-1\right)  ^{2}d_{d,k}^{\gamma}}{2\overline{P}_{k}^{2}\left(  \sum
_{j\in\mathcal{C}_{k}}1/d_{d,j}^{\gamma}\right)  }+O\left(  \overline{P}%
_{k}^{-3}\right)  . \label{AF_2h_PoutLB}%
\end{equation}
Thus, the maximum diversity of two-hop\ AF is bounded by $2$. Further, since
AF achieves a maximum diversity of $2$ with one cooperating node or relay
\cite{cap_theorems:JNL01}, allowing selection of one cooperating node with the
smallest outage, we can upper bound $P_{o}^{(k)}$ by the AF outage probability
of a relay channel with $\left\vert \mathcal{C}_{j}\right\vert =1$. Finally,
using the fact that $P_{o}^{(k)}$ for a non-orthogonal\ relay channel is at
most that for the orthogonal relay channel, we apply the high SNR\ (no CSI at
transmitters) bound developed for the latter in \cite{cap_theorems:JNL01} to
bound $P_{o}^{(k)}$ as
\begin{equation}
P_{o}^{(k)}\leq\frac{\left(  2^{2R_{k}}-1\right)  ^{2}d_{d,k}^{\gamma}%
\max_{j\in\mathcal{C}_{k}}\left(  d_{j,k}^{\gamma}+d_{d,j}^{\gamma}\right)
}{2\overline{P}_{k}^{2}}.
\end{equation}
Thus, we see that the maximum diversity achievable by a two-hop AF\ scheme in
the high SNR regime is at most $2$ and is independent of the number of
cooperating users in $\mathcal{C}_{k}$.

\subsubsection{TD-UC -- Multi-hop}

We consider an $L_{k}$-hop cooperative AF protocol where only user $k$ and
user $\pi_{k}\left(  l\right)  $, $l=1,2,\ldots,L_{k}$, transmit in the
$l^{th}$ fraction, i.e., user $\pi_{k}\left(  l\right)  $ forwards in the
fraction $\theta_{k,l}$ a scaled version of the signal it receives from user
$k$ in the first fraction. User $k$ transmits with a different codebook in the
first and second fractions. Note that $\pi_{k}\left(  1\right)  =k$ and
$\theta_{k,l}=1/L_{k}$ for all $l$. We write the received signal, $Y_{d,l}$,
at the destination in the $l^{th}$ fraction as%
\begin{equation}
Y_{d,l}=\left\{
\begin{array}
[c]{ll}%
H_{d,k}X_{k,l}+Z_{d,l} & l=1\\
H_{d,k}X_{k,l}+H_{d,\pi_{k}\left(  l\right)  }X_{\pi_{k}\left(  l\right)
,l}+Z_{d,l}^{\prime} & l=2,\ldots,L_{k}%
\end{array}
\right.  \label{AF_MH_Ydl}%
\end{equation}
where the signal transmitted by user $\pi_{k}\left(  l\right)  $ in the
$l^{th}$ fraction is $X_{\pi_{k}\left(  l\right)  ,l}=c_{\pi_{k}\left(
l\right)  }Y_{\pi_{k}\left(  l\right)  ,1}=c_{\pi_{k}\left(  l\right)
}\left(  H_{\pi_{k}\left(  l\right)  ,k}X_{k,1}+Z_{\pi_{k}\left(  l\right)
,1}\right)  ,$ and $c_{\pi_{k}\left(  l\right)  }$, $c_{s,\pi_{k}\left(
l\right)  }^{2}$, and $Z_{d,l}^{\prime}$ are given by (\ref{AF_cj}%
)-(\ref{AF_Zprime}), respectively, with $\mathcal{C}_{k}=$ $\left\{  \pi
_{k}\left(  l\right)  \right\}  $. Similar to (\ref{AF_2hop_Yd}),
(\ref{AF_MH_Ydl}) can also be written compactly as \underline{$Y$}%
$_{d}=\mathbf{H}\underline{X}_{k}+\underline{Z},$ where the $L_{k}$ entries of
\underline{$Y$}$_{d}$ and \underline{$X$}$_{k}$ are related by
(\ref{AF_MH_Ydl}) and $\mathbf{H}$ is the resulting channel gains matrix. The
destination decodes after collecting the received signals from all $L_{k}$
fractions. Choosing $X_{k,l}$, for all $l$, as independent Gaussian signals,
we have%
\begin{equation}
P_{o}^{(k)}=\Pr\left(  \log\left\vert I+\overline{P}_{k}\mathbf{HH}^{\dag
}\right\vert <L_{k}R_{k}\right)
\end{equation}
where $\mathbf{H}^{\dag}$ is the conjugate transpose of $\mathbf{H}$. We lower
bound $P_{o}^{(k)}$ with the outage probability of a $L_{k}\times1$ MIMO
channel with i.i.d. Gaussian signaling at the $L_{k}$ transmit antennas to
obtain%
\begin{equation}
P_{o}^{(k)}\geq P_{o,L_{k}\times1}=\frac{\left(  2^{R_{k}}-1\right)  ^{L_{k}%
}\prod_{l=1}^{L_{k}}d_{d,\pi_{k}\left(  l\right)  }^{\gamma}}{\left(
L_{k}!\right)  \overline{P}_{k}^{L_{k}}}+O\left(  \overline{P}_{k}^{-L_{k}%
-1}\right)  . \label{AF_MH_PoutLB}%
\end{equation}
On the other hand, one can upper bound $P_{o}^{(k)}$ by the outage probability
of an orthogonal AF protocol where user $k$ and its cooperating users transmit
on orthogonal channels, i.e., only user $\pi_{k}\left(  l\right)  $ transmits
in the fraction $\theta_{k,l}$, as developed in \cite{cap_theorems:JNL01}.
Thus, we have%
\begin{equation}
P_{out}\leq\frac{\left(  2^{L_{k}R_{k}}-1\right)  ^{L_{k}}d_{d,k}^{\gamma
}\prod_{j\in\mathcal{C}_{k}}\left(  d_{d,j}^{\gamma}+d_{j,k}^{\gamma}\right)
}{L_{k}!\overline{P}_{k}^{L_{k}}}. \label{AF_MH_PoutUB}%
\end{equation}
Comparing (\ref{AF_MH_PoutLB}) and (\ref{AF_MH_PoutUB}), we see that the
$L_{k}$-hop AF\ scheme can achieve a maximum diversity of $L_{k}$ in the high
SNR regime at the expense of user $k$ repeating the signal $L_{k}$ times.

\section{\label{Sec_4}Illustration of Results}

\begin{figure*}[tbp] \centering
{\includegraphics[
height=3.1202in,
width=4.4339in
]%
{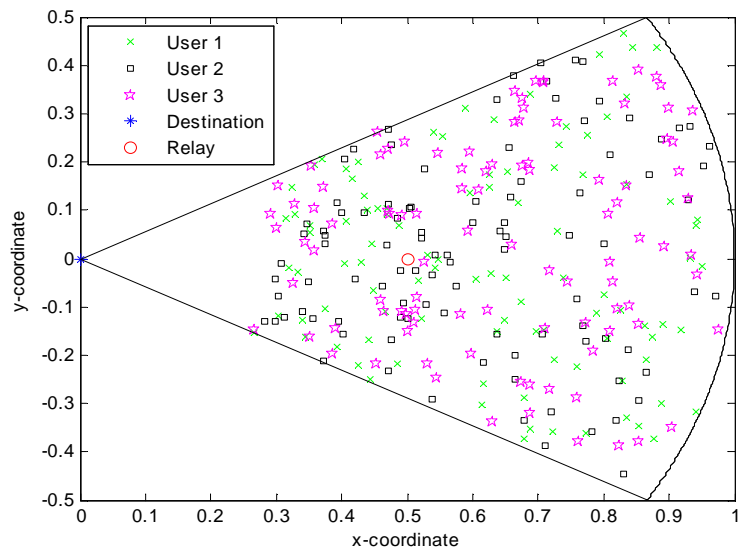}%
}%
\caption{Sector of a circle with the destination at the origin and 100 randomly chosen locations for a three-user MAC.}\label{Fig_AA_geometry}%
\end{figure*}

We consider a planar geometry with the users distributed randomly in a sector
of a circle of unit radius and angle $\pi/3$. We place the destination at the
center of the circle and place the relay at $(0.5,0)$ as shown in Fig.
\ref{Fig_AA_geometry}. The $K$ users are distributed randomly over the sector
excluding an area of radius $0.3$ around the destination. We consider $100$
such random placements and for each such random placement, we compute the
outage probabilities $P_{out}$ for the TD-RC, the TD-UC, and the TD-MAC
network as an average over the outages of all the time-duplexed users in each
network. Finally, we also average $P_{out}$ over the 100 random node
placements. We consider a three-user MAC. We assume that all three users have
the same transmit power constraint, i.e., $P_{k}=P_{1}$ for all $k$. For the
relay we choose $P_{r}=f_{r}\cdot P_{1}$ where $f_{r}\in\{0.5,1\}$. We set the
path loss exponent $\gamma=4$ and the processing factors $\eta_{k}=\delta
_{k}=\eta$ for all $k$. We plot $P_{out}$ as a function of $P_{tot}$ for
$\eta=0.01,$ $0.5,$ and $1$ thereby modeling three different regimes of
processing to transmit power ratios. We consider a symmetric transmission
rate, i.e., all users transmit at $R=0.25$ bits/channel use. We first plot
$P_{out}$ as a function of the transmit SNR $P_{1}$ in dB obtained by
normalizing $P_{1}$ by the unit variance noise. We also plot $P_{out}$ as a
function of $P_{tot}$ in dB where $P_{tot}$ is given by (\ref{Tot_pwr_costs})
and (\ref{Ptot_all_nets}). For user cooperation, we plot the outage for both
the two-hop and three-hop schemes.

\subsection{Outage Probability:\ DDF}

\begin{figure*}[tbp] \centering
{\includegraphics[
trim=0.373900in 0.027300in 0.391454in 0.058444in,
height=3.2603in,
width=6.5267in
]%
{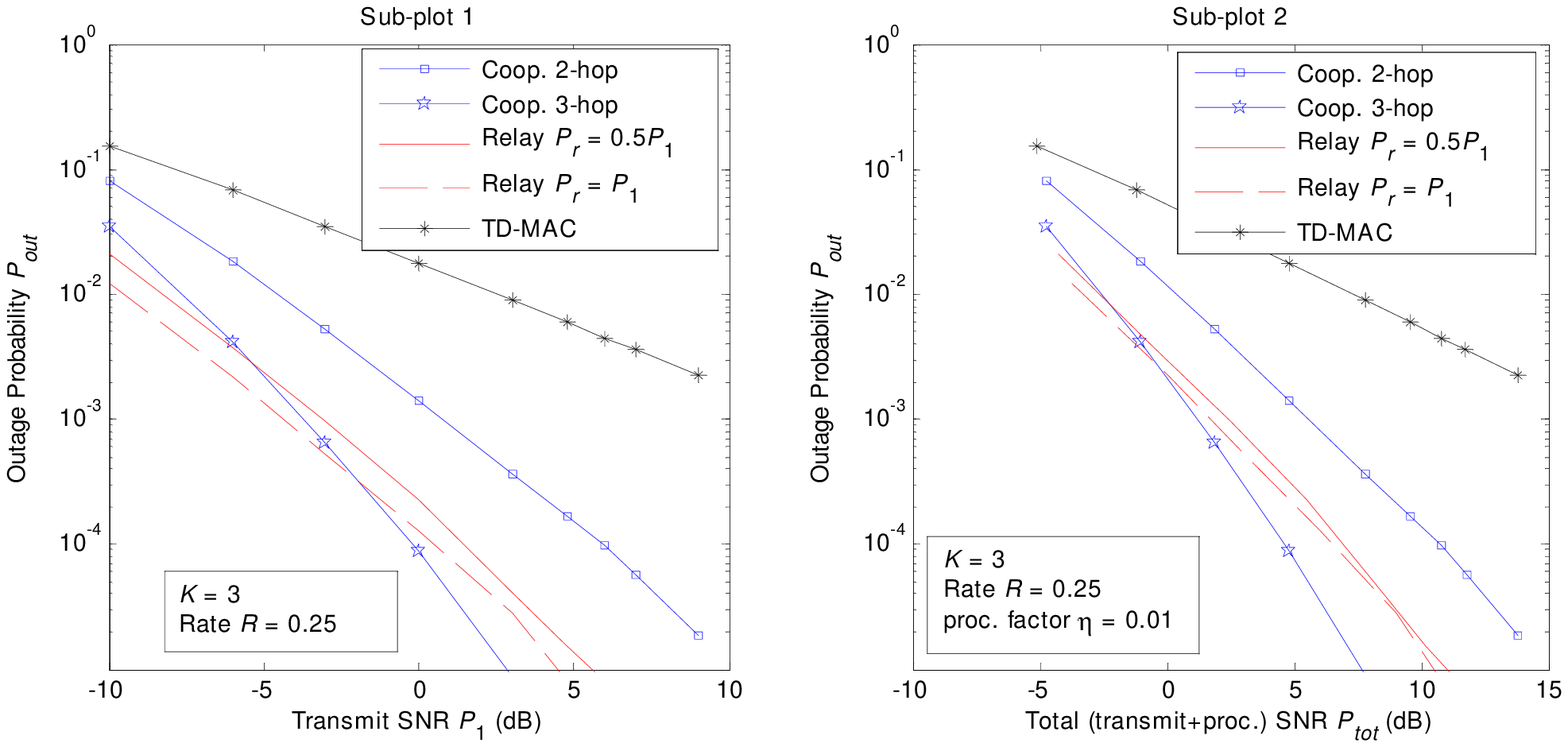}%
}%
\caption{Three user DDF outage probability $P_{out}$ vs. $P_1$ (sub-plot 1) and vs. $P_{tot}$ for $\eta=0.01$ (sub-plot 2).}\label{Fig_DFOut_K3_1}%
\end{figure*}

\begin{figure*}[tbp] \centering
{\includegraphics[
trim=0.286527in 0.027684in 0.417085in 0.059213in,
height=3.1116in,
width=6.634in
]%
{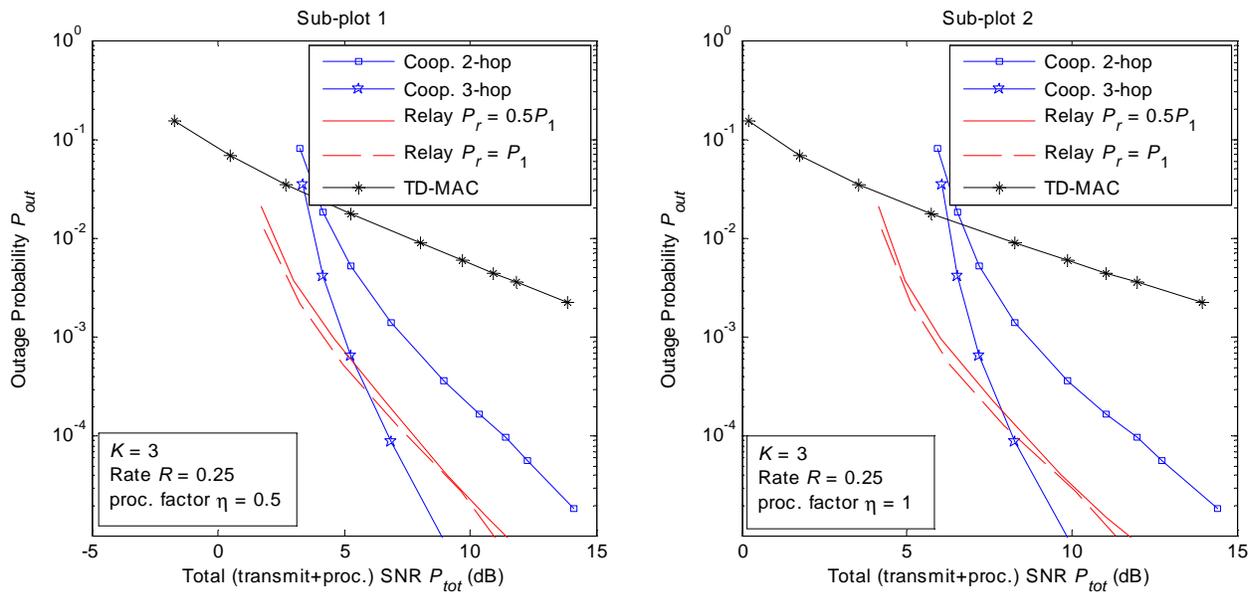}%
}%
\caption{Three user DDF outage probability $P_{out}$ vs. total transmit SNR $P_{tot}$ in dB for $\eta=0.5$ (sub-plot 1) and $\eta=1$ (sub-plot 2).}\label{Fig_DFOut_K3_2}%
\end{figure*}

We compare the outage probability of a three user MAC in Figs.
\ref{Fig_DFOut_K3_1} and \ref{Fig_DFOut_K3_2}. The plots clearly validate our
analytical results that DDF does not achieve the maximum diversity gains of
$3$ for the two hop TD-UC network (denoted Coop. 2-hop in plots). On the other
hand, the slope of $P_{out}$ for the three-hop TD-UC network, (denoted Coop.
3-hop) approaches $3$. Further, DDF for this network achieves coding gains
relative to the TD-RC network only as the SNR increases. In fact, this
difference persists even when the energy costs of cooperation are accounted
for in sub-plot 2 and Fig. \ref{Fig_DFOut_K3_2} by plotting $P_{out}$ as a
function of $P_{tot}$. This difference in SNR\ gains between user and relay
cooperation is due to the fact that user cooperation increases spatial
diversity at the expense of requiring users to share their power for
cooperative transmissions. Observe that with increasing $\eta$, the outage
curves are translated to the right. In fact, for a fixed $R$, the processing
costs increase with increasing $\eta$, and thus, we expect the SNR gains from
cooperation to diminish relative to TD-MAC, particularly in the lower SNR
regimes of interest. This is demonstrated in Fig. \ref{Fig_DFOut_K3_2}.

\subsection{Outage Probability:\ AF}

\begin{figure*}[tbp] \centering
{\includegraphics[
trim=0.171300in 0.041492in 0.216747in 0.063984in,
height=3.141in,
width=6.5925in
]%
{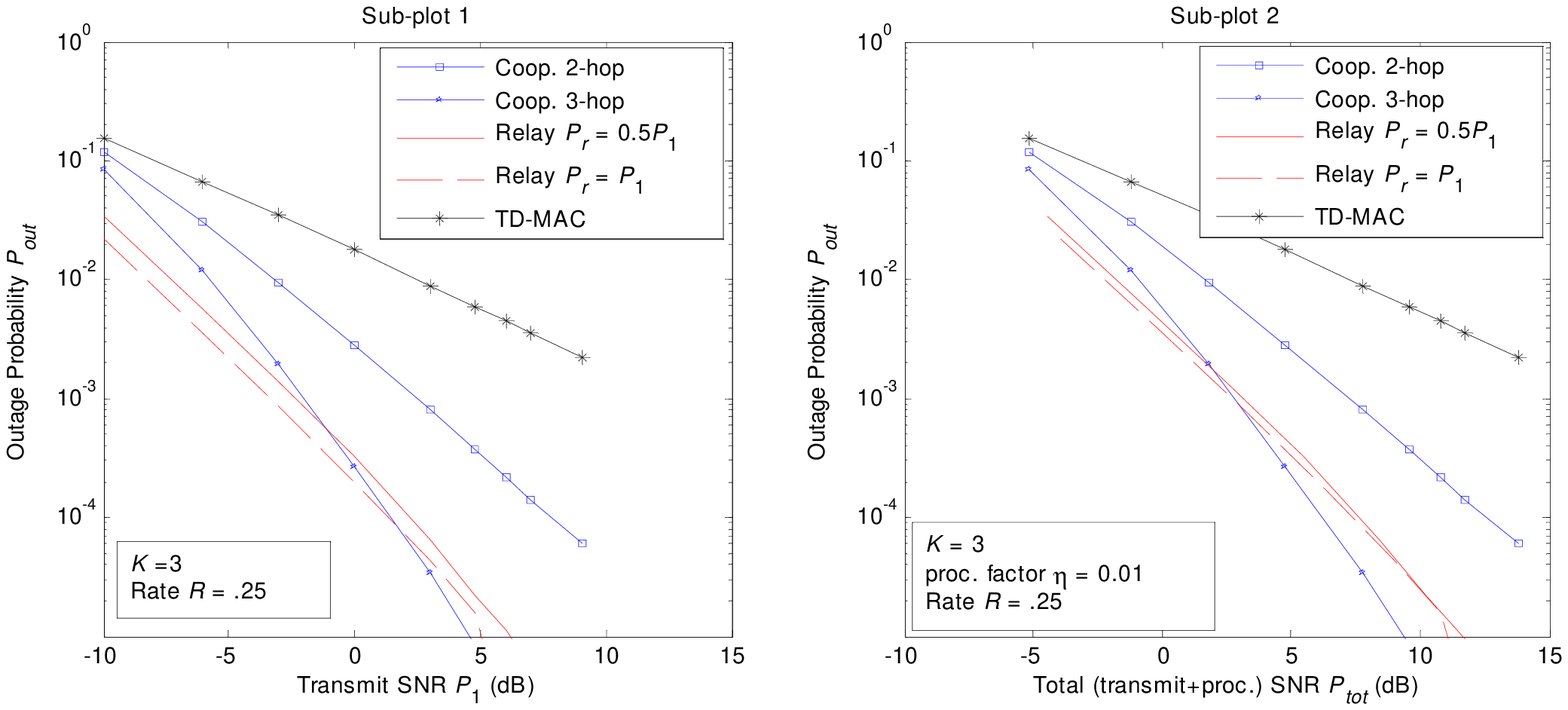}%
}%
\caption{Three user AF outage probability $P_{out}$ vs. $P_1$ (sub-plot 1) and $P_{tot}$ for $\eta=0.01$ (sub-plot 2).}\label{Fig_AFOut_K3_1}%
\end{figure*}

\begin{figure*}[tbp] \centering
{\includegraphics[
trim=0.172617in 0.028270in 0.217305in 0.043760in,
height=3.1496in,
width=6.6556in
]%
{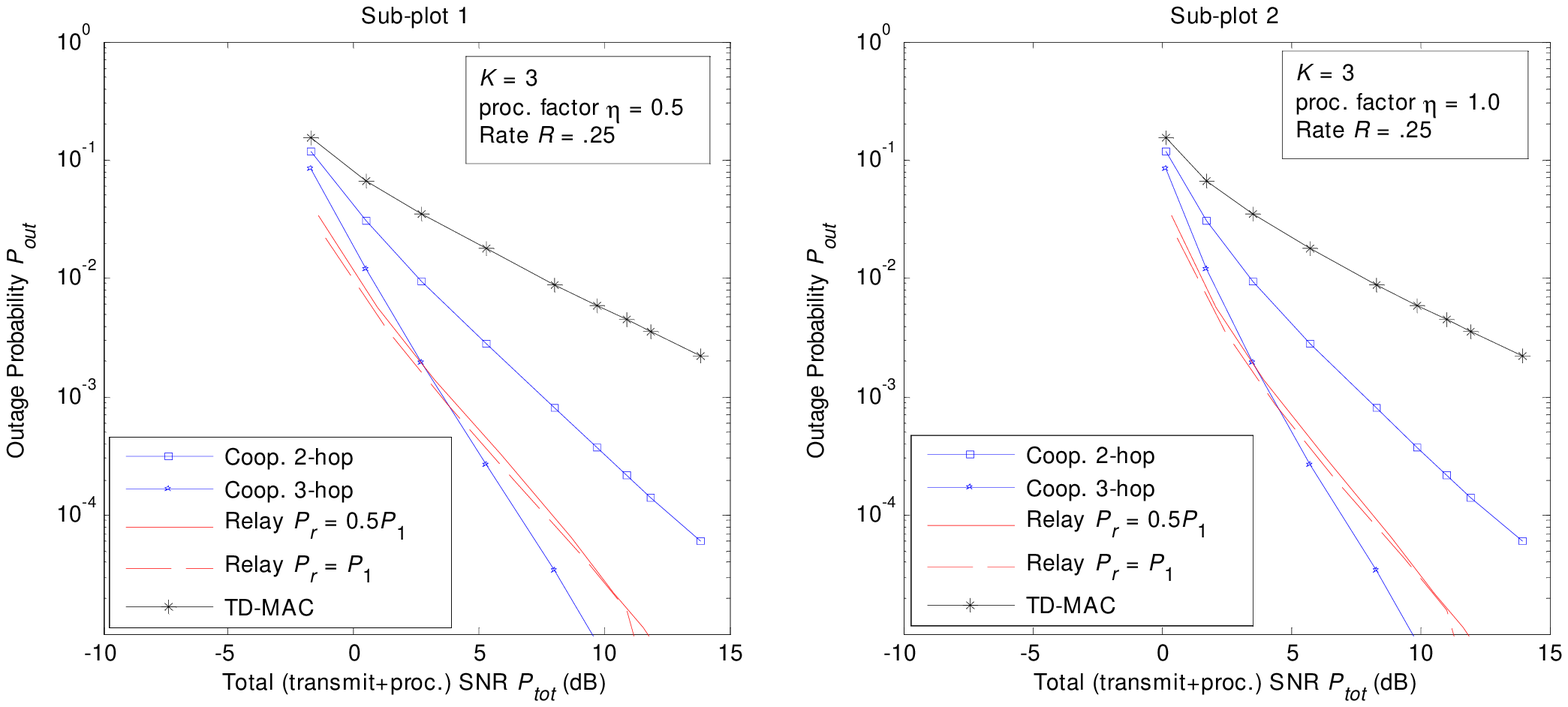}%
}%
\caption{Three user AF outage probability $P_{out}$ vs. $P_{tot}$ for $\eta=0.5$ (sub-plot 1) and $\eta=1$ (sub-plot 2).}\label{Fig_AFOut_K3_2}%
\end{figure*}

In Figs. \ref{Fig_AFOut_K3_1} and \ref{Fig_AFOut_K3_2} we plot the two user
AF\ outage probability for all three networks. As predicted, we see that both
TD-RC and TD-UC networks achieve a maximum diversity of $2$ for the two-hop
scheme. The three-hop scheme for TD-UC achieves a maximum diversity
approaching $3$. However, it achieves coding gains relative to the relay
network only as the SNR increases. These gains are a result of the model
chosen for the processing power (only model costs of encoding and decoding)
and the choice of $P_{k,0}^{proc}=0$ for all $k$ for the purposes of
illustration. In general, $P_{k,0}^{proc}>0$ since it models protocol and
device overhead including front-end processing and amplification costs, and
thus, the total processing power will scale proportionate to the number of
users that a node relays for.

The numerical analysis can be extended to arbitrary relay positions
\cite[Chap. 4]{cap_theorems:LalithaSankar}. In general, the choice of relay
position is a tradeoff between cooperating with as many users as possible and
being, on average, closer than the users are to the destination. To this end,
fixing the relay at the symmetric location of $(0.5,0)$ is a reasonable tradeoff.

\section{\label{Sec_5}Concluding Remarks}

We compared the outage performance of user and relay cooperation in a
time-duplexed multiaccess network using the total transmit and processing
power as a cost metric for the comparison. We developed a model for processing
power costs as a function of the transmitted rate. We developed a two-hop
cooperation scheme for both the relay and user cooperative network. We also
presented a multi-hop scheme for the user cooperative network for the case of
multiple cooperating users. We presented geometry-inclusive upper and lower
bounds on the outage probability of DDF and AF to facilitate comparisons of
diversity and coding gains achieved by the two cooperative approaches.\ We
showed that the TD-RC network achieves a maximum diversity of $2$ for both DDF
and AF. We also showed that under a two-hop transmission scheme, a $K$-user
TD-UC network achieves a $K$-fold diversity gain with DDF\ only when the
cooperating users are physically proximal and achieves a maximum diversity of
$2$ with AF. On the other hand, for a $K$-hop transmission scheme, the TD-UC
network achieves a maximum diversity of $K$ for both DDF and AF. Using
area-averaged numerical results that account for the costs of cooperation, we
demonstrated that the TD-RC network achieves SNR gains that either diminish or
completely eliminate the diversity advantage of the TD-UC network in SNR
ranges of interest. Besides a fixed relay position, this difference is due to
the fact that user cooperation results in a tradeoff between diversity and SNR
gains as a result of sharing limited power resources between the users.

In conclusion, we see that user cooperation is desirable only if the
processing costs associated with achieving the maximum diversity gains are not
prohibitive, i.e., in the regime where user cooperation achieves positive
coding gains relative to the relay cooperative and non-cooperative networks.
The simple processing cost model presented here captures the effect of
transmit rate on processing power. One can also tailor this model to
explicitly include delay, complexity, and device-specific processing costs.%

\appendices

\section{\label{Comp_App0}Distribution of Weighted Sum of Exponential Random
Variables}

Consider a collection of i.i.d. unit mean exponential random variables $E_{l}%
$, $l=1,2,\ldots,L$. We denote a weighted sum of $E_{l}$, for all $l$, as
$H=\sum_{l=1}^{L}c_{l}E_{l}$ where $c_{l}>0$ and $c_{m}\not =c_{k}$ for all
$l$ and $m\not =k$. The following lemma summarizes the probability
distribution of $H$ \cite[p. 11]{cap_theorems:Cox01}.

\begin{lemma}
[{\cite[p. 11]{cap_theorems:Cox01}}]\label{Lemma_Hypo}The random variable $H$
has a distribution given as
\begin{equation}
p_{H}\left(  h\right)  =\left\{
\begin{array}
[c]{ll}%
\sum_{l=1}^{L}\frac{C_{l}}{c_{l}}e^{-h/c_{l}} & h\geq0\\
0 & \text{otherwise}%
\end{array}
\right.  \label{A1_dist}%
\end{equation}
where the constants $C_{l}$, for all $l$, are
\begin{equation}
C_{l}=\left\{
\begin{array}
[c]{ll}%
1 & L=1\\
\frac{\left(  -c_{l}\right)  ^{L-1}}{\prod_{j=1,j\not =l}^{L}\left(
c_{j}-c_{l}\right)  } & L>1.
\end{array}
\right.  \label{A1_coeff}%
\end{equation}
The cumulative distribution function of $H$ is
\begin{equation}
F_{H}\left(  \eta\right)  =\sum_{l=1}^{L}C_{l}\left(  1-e^{-\eta/c_{l}%
}\right)  \label{A1_Fexp}%
\end{equation}
such that the first non-zero term in the Taylor series expansion of
$F_{H}\left(  \eta\right)  $ about $\eta=0$ is $\eta^{L}\left/  L!\left(
\prod_{l=1}^{L}c_{l}\right)  \right.  $.
\end{lemma}

\section{\label{Comp_App2}DDF\ Outage Bounds}

\subsection{Two-Hop Relay Cooperative Network}

For a DDF relay, the listen fraction is the random variable (see \cite[(13),
pp. 4157]{cap_theorems:AGS01})%
\begin{equation}
\Theta_{k}=\min\left(  1,R_{k}\left/  \log\left(  1+\frac{\left\vert
A_{r,k}\right\vert ^{2}\overline{P}_{k}}{d_{r,k}^{\gamma}}\right)  \right.
\right)  . \label{A2_theta_def}%
\end{equation}
$\Theta_{k}$ is a mixed (discrete and continuous) random variable with a
cumulative distribution function (CDF) given as%
\begin{equation}
F_{\Theta_{k}}^{(r)}(\theta_{k})=\left\{
\begin{array}
[c]{ll}%
0 & \theta_{k}\leq0\\
\exp\left[  -\frac{(2^{R_{k}/\theta_{k}}-1)d_{r,k}^{\gamma}}{\overline{P}_{k}%
}\right]  & 0<\theta_{k}<1\\
1 & \theta_{k}=1.
\end{array}
\right.  \label{A2_CDF_Theta}%
\end{equation}
The mutual information collected at the destination over both the
\textit{listen} and \textit{transmit} fractions is (see \cite[Appendix
D]{cap_theorems:AGS01})%
\begin{equation}
I_{2}^{DF}=\Theta_{k}G_{1}+\overline{\Theta}_{k}G_{2} \label{A2_I2r}%
\end{equation}
where $\overline{\Theta}_{k}=\left(  1-\Theta_{k}\right)  $, $\overline{P}%
_{k}=KP_{k}$, $\overline{P}_{r}=P_{r}/\overline{\Theta}_{k}$, and
\begin{align}
G_{1}  &  =C\left(  \left\vert H_{d,k}\right\vert ^{2}\overline{P}_{k}\right)
\\
G_{2}  &  =C\left(  \left\vert H_{d,k}\right\vert ^{2}\overline{P}%
_{k}+\left\vert H_{d,r}\right\vert ^{2}\overline{P}_{r}\right)  .
\end{align}
The outage probability for user $k$ transmitting at a fixed rate $R_{k}$ is
then given as%
\begin{equation}
P_{o}^{(k)}=\Pr\left(  I_{2}^{DF}<R_{k}\right)  \label{A2_Pout}%
\end{equation}
From (\ref{A2_theta_def}), $\Theta_{k}=0$ only for $d_{r,k}=0$, i.e., only
when user $k$ and the relay are co-located, and for this case (\ref{A2_Pout})
simplifies to the outage probability of a $2\times1$ MIMO channel given as%

\begin{equation}
P_{o,2\times1}=\Pr\left(  C\left(  \frac{\left\vert A_{d,k}\right\vert
^{2}\overline{P}_{k}}{d_{d,k}^{\gamma}}+\frac{\left\vert A_{d,r}\right\vert
^{2}P_{r}}{d_{d,r}^{\gamma}}\right)  <R_{k}\right)  .
\end{equation}
Let $P_{r}$ and $\overline{P}_{k}$ scale such that $P_{r}/\overline{P}%
_{k}=\lambda$ is a positive constant. Using (\ref{A1_Fexp}), we have
\begin{equation}
P_{o,2\times1}=\frac{(2^{R_{k}}-1)^{2}d_{d,k}^{\gamma}d_{d,r}^{\gamma}%
}{2\lambda\overline{P}_{k}^{2}}+O\left(  \overline{P}_{k}^{-3}\right)  .
\label{A2_Pout_LB}%
\end{equation}
$P_{o,2\times1}$ is a lower bound on $P_{o}^{(k)}$ because $G_{2}\geq G_{1}$.
On the other hand, for any $\theta_{k}$ in (\ref{A2_I2r}), $P_{o}^{(k)}%
(\theta_{k})$ can be upper bounded as
\begin{align}
P_{o}^{(k)}(\theta_{k})  &  \leq\Pr\left(  \theta_{k}G_{1}<R_{k}\right)
=P_{o,1}^{(k)}(\theta_{k})\label{A2_PUB1}\\
P_{o}^{(k)}(\theta_{k})  &  \leq\Pr\left(  \overline{\theta}_{k}G_{2}%
<R_{k}\right)  =P_{o,2}^{(k)}(\theta_{k}) \label{A2_PUB2}%
\end{align}
Thus, we have%
\begin{equation}
P_{o}^{(k)}=\mathbb{E}P_{o}^{(k)}(\Theta_{k})\leq\mathbb{E}\min(P_{o,1}%
^{(k)}(\Theta_{k}),P_{o,2}^{(k)}(\Theta_{k}))=P_{UB}^{(k)} \label{A2_Po_PUB}%
\end{equation}
Let%
\begin{equation}%
\begin{array}
[c]{ccc}%
\eta=2^{R_{k}/\overline{\theta}_{k}}-1, & c_{1}=\frac{\overline{P}_{k}%
}{d_{d,k}^{\gamma}}, & c_{2}=\frac{\overline{P}_{r}}{d_{d,r}^{\gamma}}.
\end{array}
\end{equation}
From (\ref{A1_coeff}), we have $C_{1}=c_{1}\left/  \left(  c_{1}-c_{2}\right)
\right.  $ and $C_{2}=c_{2}\left/  \left(  c_{2}-c_{1}\right)  \right.  $.

Using Lemma \ref{Lemma_Hypo}, we can expand $P_{o,1}^{(k)}(\theta_{k})$ and
$P_{o,2}^{(k)}(\theta_{k})$ in (\ref{A2_Po_PUB}) as%
\begin{equation}
P_{o,1}^{(k)}(\theta_{k})=\Pr\left(  G_{1}<\frac{R_{k}}{\theta_{k}}\right)
=1-\exp\left[  \frac{-(2^{R_{k}/\theta_{k}}-1)d_{d,k}^{\gamma}}{\overline
{P}_{k}}\right]  \leq\frac{(2^{R_{k}/\theta_{k}}-1)d_{d,k}^{\gamma}}%
{\overline{P}_{k}} \label{A2PU0}%
\end{equation}%
\begin{equation}
P_{o,2}^{(k)}(\theta_{k})=\Pr\left(  G_{2}<\frac{R_{k}}{\overline{\theta}_{k}%
}\right)  =\sum_{l=1}^{2}C_{l}\left(  1-e^{-\eta/c_{l}}\right)  =\frac
{(2^{R_{k}/\overline{\theta}_{k}}-1)^{2}\overline{\theta}_{k}d_{d,k}^{\gamma
}d_{d,r}^{\gamma}}{2\lambda\overline{P}_{k}^{2}}+O\left(  \overline{P}%
_{k}^{-3}\right)  \label{A2_PUB}%
\end{equation}
where the bound in (\ref{A2_PUB}) follows from expanding and simplifying the
exponential functions. From (\ref{A2_PUB}), we see that for a fixed
$\overline{P}_{k}$ and $d_{j,k}$ for all $j,k$, the minimum in
(\ref{A2_Po_PUB}) is dominated by $P_{o,2}^{(k)}(\theta_{k})$ for small
$\theta_{k}$ and by $P_{o,1}^{(k)}(\theta_{k})$ as $\theta_{k}$ approaches
$1$. Finally, we have $P_{o,2\times1}=P_{o,2}^{(k)}(\theta_{k}=0)$.

In general, $P_{UB}^{(k)}$ is not easy to evaluate analytically. Since we are
interested in the achievable diversity, we develop a bound on $P_{UB}^{(k)}$
for a fixed $R_{k}$. We have, for any $\theta_{k}^{\ast}$, $0<\theta_{k}%
^{\ast}<1$,%
\begin{align}
P_{UB}^{(k)}  &  =\int_{0}^{1}P_{\Theta_{k}}\left(  \theta_{k}\right)
\min\left(  P_{o,1}^{(k)}(\theta_{k}),P_{o,2}^{(k)}(\theta_{k})\right)
d\theta_{k}\label{A2PUB_B0}\\
&  \leq\int_{0}^{\theta_{k}^{\ast}}P_{\Theta_{k}}\left(  \theta_{k}\right)
P_{o,2}^{(k)}(\theta_{k})d\theta_{k}+\int_{\theta_{k}^{\ast}}^{1}P_{\Theta
_{k}}\left(  \theta_{k}\right)  P_{o,1}^{(k)}(\theta_{k})d\theta
_{k}\label{A2PUB_B1}\\
&  \leq F_{\Theta_{k}}\left(  \theta_{k}^{\ast}\right)  P_{o,2}^{(k)}%
(\theta_{k}^{\ast})+\left(  1-F_{\Theta_{k}}\left(  \theta_{k}^{\ast}\right)
\right)  P_{o,1}^{(k)}(\theta_{k}^{\ast})\label{A2_PUB_B2}\\
&  \leq P_{o,2}^{(k)}(\theta_{k}^{\ast})+\frac{(2^{R_{k}/\theta_{k}^{\ast}%
}-1)d_{r,k}^{\gamma}}{\overline{P}_{k}}\cdot P_{o,1}^{(k)}(\theta_{k}^{\ast
})\label{A2_PUB_B3}\\
&  \leq\left[  \frac{(2^{R_{k}/\overline{\theta}_{k}^{\ast}}-1)^{2}%
\overline{\theta}_{k}^{\ast}}{(2^{R_{k}}-1)^{2}}+\frac{2d_{r,k}^{\gamma
}(2^{R_{k}/\theta_{k}^{\ast}}-1)^{2}\lambda}{d_{d,r}^{\gamma}(2^{R_{k}}%
-1)^{2}}\right]  \cdot\frac{(2^{R_{k}}-1)^{2}d_{d,k}^{\gamma}d_{d,r}^{\gamma}%
}{2\lambda\overline{P}_{k}^{2}}+O\left(  \overline{P}_{k}^{-3}\right)
\label{A2_PkUB_final}%
\end{align}
where the equality in (\ref{A2PUB_B1}) holds when $P_{o,2}^{(k)}(\theta
_{k})<P_{o,1}^{(k)}(\theta_{k})$ for $\theta_{k}<\theta_{k}^{\ast}$ and
vice-versa, and (\ref{A2_PUB_B2}) follows because $P_{o,1}^{(k)}(\theta_{k})$
and $P_{o,2}^{(k)}(\theta_{k})$ decrease and increase, respectively, with
$\theta_{k}$ and (\ref{A2_PUB_B3}) follows from using (\ref{A2_CDF_Theta}) to
bound $1-F_{\Theta_{k}}\left(  \theta_{k}^{\ast}\right)  $. Finally, we note
that for any fixed $0<\theta_{k}^{\ast}<1$, for fixed inter-node distances,
the term in square brackets in (\ref{A2_PkUB_final}) is a multiplicative
constant separating the upper bound (\ref{A2_PkUB_final}) and the lower bound
(\ref{A2_Pout_LB}) on $P_{o}^{\left(  k\right)  }$.

\subsection{Two-hop User Cooperative Network}

The above analysis extends to the two-hop TD-UC network. Recall that a DDF
cooperating node remains in the \textit{listen} mode until it successfully
decodes its received signal from the source. Thus, for the two-hop TD-UC
network, the \textit{listen} fraction for each cooperating node $j$, for all
$j\in\mathcal{C}_{k}$, is given by (\ref{A2_theta_def}) with the substition
$r=j$. Further, since the \textit{listen }fraction $\Theta_{k}$ is now the
largest among all $j$, from (\ref{A2_theta_def}) we have%
\begin{equation}
\Theta_{k}=\min\left(  1,\max\nolimits_{j\in\mathcal{C}_{k}}\left\{
R_{k}\left/  C\left(  \left.  \left\vert A_{j,k}\right\vert ^{2}\overline
{P}_{k}\right/  d_{j,k}^{\gamma}\right)  \right.  \right\}  \right)
\label{A3_thetk_def}%
\end{equation}
where the transmit power $\overline{P}_{k}$, for all $k\in\mathcal{K}$,
satisfies (\ref{MAC_pwr_constrant}) and is given by (\ref{UC_Khop_Pwr}). Let
$F_{\Theta_{k}}^{(j)}(\theta_{k})$ be the CDF $F_{\Theta_{k}}^{(r)}(\theta
_{k})$ in (\ref{A2_CDF_Theta}) with the index $r$ replaced by $j$. From the
independence of $A_{j,k}$ for all $j\in\mathcal{C}_{k}$, the CDF of
$\Theta_{k}$ is
\begin{equation}
F_{\Theta_{k}}(\theta_{k})=\prod_{j\in\mathcal{C}_{k}}F_{\Theta_{k}}%
^{(j)}(\theta_{k})=\left.  F_{\Theta_{k}}^{(r)}(\theta_{k})\right\vert
_{d_{r,k}=%
{\textstyle\sum_{j\in C_{k}}}
d_{j,k}^{\gamma}}.
\end{equation}
The destination collects information from the transmissions of user $k$ and
all its cooperating nodes in $\mathcal{C}_{k}$ over both the \textit{transmit}
and \textit{listen} fractions. The resulting mutual information achieved by
user $k$ at the destination is (see \cite{cap_theorems:GKR01})%
\begin{equation}
I_{2,DF}^{c}(\Theta_{k})=\Theta_{k}G_{1}+\overline{\Theta}_{k}G_{2}
\label{A3_I2_c}%
\end{equation}
where $\overline{\Theta}_{k}=1-\Theta_{k}$ and
\begin{align}
G_{1}  &  =C\left(  \left\vert H_{d,k}\right\vert ^{2}\overline{P}_{k}\right)
\\
G_{2}  &  =C\left(  \left\vert H_{d,k}\right\vert ^{2}\overline{P}_{k}%
+\sum_{j\in\mathcal{C}_{k}}\left\vert H_{d,j}\right\vert ^{2}\frac
{\overline{P}_{j}}{\overline{\Theta}_{k}}\right)  .
\end{align}
The DDF outage probability for user $k$ transmitting at a fixed rate $R_{k}$
in a two-hop TD-UC network is thus given as
\begin{equation}
P_{o}^{(k)}=\Pr\left(  I_{2,DF}^{c}<R_{k}\right)  .
\end{equation}
From (\ref{A3_thetk_def}), $\Theta_{k}=0$ only if $d_{j,k}=0$ for all
$j\in\mathcal{C}_{k}$.

From (\ref{A3_I2_c}), we can lower bound $P_{o}^{(k)}$ by the outage
probability, $P_{o,L_{k}\times1}$, of a $L_{k}\times1$ distributed MIMO
channel given as%

\begin{equation}
P_{o,L_{k}\times1}=\Pr\left(  C\left(  \frac{\left\vert A_{d,k}\right\vert
^{2}\overline{P}_{k}}{d_{d,k}^{\gamma}}+\sum_{j\in\mathcal{C}_{k}}%
\frac{\left\vert A_{d,j}\right\vert ^{2}\overline{P}_{j}}{d_{d,j}^{\gamma}%
}\right)  <R_{k}\right)  .
\end{equation}
We enumerate the $\left(  L_{k}-1\right)  $ cooperative nodes in
$\mathcal{C}_{k}$ as $l=2,3,\ldots,L_{k}$, and write $\mathcal{S}_{k}=\left\{
k\right\}  \cup\mathcal{C}_{k}$. Using (\ref{A1_Fexp}), and scaling
$\overline{P}_{j}$ and $\overline{P}_{k}$ such that $\overline{P}%
_{j}/\overline{P}_{k}=$ $\lambda_{j}$ is a constant, for all $j$, we have
\begin{equation}
P_{o,L_{k}\times1}=\frac{(2^{R_{k}}-1)^{L_{k}}d_{d,k}^{\gamma}}{\left(
L_{k}!\right)  \overline{P}_{k}^{L_{k}}}\cdot\left(  \prod_{j\in
\mathcal{S}_{k}}\frac{d_{d,j}^{\gamma}}{\lambda_{j}}\right)  +O\left(
\overline{P}_{k}^{-L_{k}-1}\right)  . \label{A3_Po_MIMO}%
\end{equation}
Let
\begin{equation}%
\begin{array}
[c]{ccc}%
\eta=2^{R_{k}/\overline{\theta}_{k}}-1, & c_{1}=\frac{\overline{P}_{k}%
}{d_{d,k}^{\gamma}}, &
\begin{array}
[c]{cc}%
c_{l}=\frac{\overline{P}_{l}}{d_{d,l}^{\gamma}\overline{\theta}_{k}}, &
l=2,3,\ldots,L_{k}%
\end{array}
\end{array}
\end{equation}
where the $\overline{\theta}_{k}$ in $c_{l}$ is due to the definition of
$\overline{P}_{l}$ in (\ref{UC_Khop_Pwr}). The $C_{l}$, for all $l=1,2,\ldots
,L_{k}$, are given by (\ref{A1_coeff}). For a fixed $R_{k}$, we upper bound
$P_{o}^{(k)}$ using (\ref{A2_PUB1})-(\ref{A2_PUB2}) as
\begin{equation}
P_{o}^{(k)}=\mathbb{E}P_{o}^{(k)}(\Theta_{k})\leq\mathbb{E}\min(P_{o,1}%
^{(k)}(\Theta_{k}),P_{o,2}^{(k)}(\Theta_{k}))=P_{UB}^{\left(  k\right)  }.
\label{A3_Po_UB}%
\end{equation}
We upper bound $P_{o,1}^{(k)}(\theta_{k})$ using (\ref{A2PU0}) and compute
\begin{equation}
P_{o,2}^{(k)}(\theta_{k})=\frac{\left(  2^{R_{k}/\overline{\theta}_{k}%
}-1\right)  ^{L_{k}}\left(  \overline{\theta}_{k}\right)  ^{L_{k}-1}}{\left(
L_{k}!\right)  \left(  \overline{P}_{k}\right)  ^{L_{k}}}\left(  \prod
_{j\in\mathcal{S}_{k}}\frac{d_{d,j}^{\gamma}}{\lambda_{j}}\right)  +O\left(
\overline{P}_{k}^{-L_{k}-1}\right)  . \label{A3_Po2_finalexp}%
\end{equation}
Analogous to the steps in (\ref{A2PUB_B0})-(\ref{A2_PkUB_final}) for the TD-RC
case, we have (see (\ref{A3_Po_UB})), for any $\theta_{k}^{\ast}$,
$0<\theta_{k}^{\ast}<1$,%
\begin{align}
P_{UB}^{(k)}  &  \leq F_{\Theta_{k}}\left(  \theta_{k}^{\ast}\right)
P_{o,2}^{(k)}(\theta_{k}^{\ast})+\left(  1-F_{\Theta_{k}}\left(  \theta
_{k}^{\ast}\right)  \right)  P_{o,1}^{(k)}(\theta_{k}^{\ast})\\
&  \leq P_{o,2}^{(k)}(\theta_{k}^{\ast})+\frac{(2^{R_{k}/\theta_{k}^{\ast}%
}-1)\left(  \sum_{j\in C_{k}}d_{j,k}^{\gamma}\right)  }{\overline{P}_{k}}\cdot
P_{o,1}^{(k)}(\theta_{k}^{\ast})\label{A3P1}\\
&  =\left[  \frac{(2^{R_{k}/\overline{\theta}_{k}^{\ast}}-1)^{L_{k}}\left(
\overline{\theta}_{k}^{\ast}\right)  ^{L_{k}-1}}{(2^{R_{k}}-1)^{L_{k}}}%
+\frac{\left(  L_{k}!\right)  \left(  \sum\limits_{j\in\mathcal{C}_{k}}%
d_{j,k}^{\gamma}\right)  \overline{P}_{k}^{L_{k}-2}(2^{R_{k}/\theta_{k}^{\ast
}}-1)^{2}}{\left(  \prod\nolimits_{j\in\mathcal{C}_{k}}d_{d,j}^{\gamma
}/\lambda_{j}\right)  (2^{R_{k}}-1)^{L_{k}}}\right]  \cdot\nonumber\\
&  \text{ \ \ }\left[  \frac{(2^{R_{k}}-1)^{L_{k}}}{\left(  L_{k}!\right)
\overline{P}_{k}^{L_{k}}}\left(  \prod_{j\in\mathcal{S}_{k}}\frac
{d_{d,j}^{\gamma}}{\lambda_{j}}\right)  \right]  +O\left(  \overline{P}%
_{k}^{-L_{k}-1}\right)  . \label{A3Pub}%
\end{align}

\section{\label{Comp_App4}Multi-hop Cooperative Network -- DDF Outage
Analysis}

The DDF outage probability of user $k$ transmitting at a fixed rate $R_{k}$ in
a multi-hop user cooperative network is
\begin{equation}
P_{o}^{(k)}=\Pr\left(  I_{2,DF}^{c}<R_{k}\right)  \label{A4_Pout}%
\end{equation}
where
\begin{equation}
I_{2,DF}^{c}(\Theta_{k})=\sum_{l=1}^{L_{k}}\Theta_{k,l}G_{l}\text{.}
\label{A4_I2_c}%
\end{equation}
The function $G_{l}$ is given by%
\begin{equation}%
\begin{array}
[c]{cc}%
G_{l}=C\left(  \sum_{j=1}^{l}\left\vert H_{d,\pi_{k}(j)}\right\vert ^{2}%
\frac{\overline{P}_{\pi_{k}(j)}}{\overline{\Theta}_{k,j}}\right)  &
l=1,2,\ldots L_{k},
\end{array}
\end{equation}
where $\overline{P}_{k}$ is given by (\ref{UC_Khop_Pwr}) and
\begin{align}
&
\begin{array}
[c]{cc}%
\Theta_{k,l}^{sum}\overset{\vartriangle}{=}\sum_{j=1}^{l-1}\Theta_{k,j} &
\text{for }l=1,2,\ldots,L_{k}%
\end{array}
\label{A4_alltheta_defs}\\
&
\begin{array}
[c]{cc}%
\overline{\Theta}_{k,l}^{sum}\overset{\vartriangle}{=}1-\Theta_{k,l}^{sum} &
\end{array}
\end{align}
with $\overline{\Theta}_{k,L_{k}}^{sum}=\Theta_{k,L_{k}}$ and $\Theta
_{k,-1}=0$ such that $\overline{\Theta}_{k,1}^{sum}=1$. Recall that $\pi
_{k}\left(  \cdot\right)  $ is a permutation on $\mathcal{C}_{k}$ such that
user $\pi_{k}\left(  l\right)  $ begins its transmissions in the fraction
$\Theta_{k,l}$, for all $l=2,3,\ldots,L_{k}$. Furthermore, $\pi_{k}\left(
1\right)  =k$ and we write $\pi_{k}\left(  i:j\right)  =\{\pi_{k}(i),$
$\pi_{k}(i+1),$ $\ldots,$ $\pi_{k}(j)\}$.

We write $\underline{\Theta}_{k}$ to denote a $\left(  L_{k}-1\right)
$-length random vector with entries $\Theta_{k,l}$, $l=1,2,\ldots,L_{k}-1$,
and $\lambda_{\pi_{k}\left(  j\right)  }=\overline{P}_{\pi_{k}\left(
j\right)  }/\overline{P}_{k}$ for all $\pi_{k}\left(  j\right)  \in
\mathcal{C}_{k}$. Further, we write $\underline{\Theta}_{k}^{\left(  l\right)
}$ to denote the vector of the first $l$ entries of $\underline{\Theta}_{k}$.
The fraction $\Theta_{k,l}$, $l=1,2,\ldots,L_{k}-1$, is the smallest value
such that at least one new node, denoted as $\pi_{k}\left(  l+1\right)  $,
decodes the message from user $k$. The analysis for this problem seems
difficult; so we replace it by analyzing a simpler strategy where node
$\pi_{k}\left(  l+1\right)  $ collects energy only in fraction $\Theta_{k,l}$
from the transmissions of user $k$ as well as the users in $\pi_{k}\left(
1:l\right)  $. For this strategy, we have
\begin{equation}
\Theta_{k,l}=\min\left\{  \overline{\Theta}_{k,l}^{sum},\min_{\pi_{k}\left(
l+1\right)  \in\mathcal{C}_{k}\backslash\pi_{k}\left(  1:l\right)  }%
\frac{R_{k}}{C\left(  \sum_{m=1}^{l}\left.  \left\vert A_{\pi_{k}\left(
l+1\right)  ,\pi_{k}(m)}\right\vert ^{2}\overline{P}_{\pi_{k}(m)}\right/
d_{\pi_{k}\left(  l+1\right)  ,\pi_{k}(m)}^{\gamma}\right)  }\right\}  .
\label{A4_theta_kl_bound}%
\end{equation}
Applying Lemma \ref{Lemma_Hypo}, the CDF of $\Theta_{k,l}$ conditioned on
\underline{$\Theta$}$_{k}^{l-1}=\underline{\theta}_{k}^{l-1}$ simplifies to
\begin{equation}
F_{\Theta_{k,l}|\underline{\Theta}_{k,1}^{l-1}}(\theta_{k,l}|\underline
{\theta}_{k}^{l-1})=\left\{
\begin{array}
[c]{ll}%
0 & \theta_{k,l}\leq0\\
1-\prod\nolimits_{j\in\mathcal{C}_{k}\backslash\pi_{k}\left(  2:l\right)
}\left[  F_{H_{j,l}^{sum}}(2^{R_{k}/\theta_{k,l}}-1)\right]  & 0<\theta
_{k,l}<\overline{\theta}_{k,l}^{sum}\\
1 & \theta_{k,l}=\overline{\theta}_{k,l}^{sum}.
\end{array}
\right.  \label{A4_Fthethak}%
\end{equation}
where from (\ref{A4_theta_kl_bound}), $H_{j,l}^{sum}\overset{\vartriangle}%
{=}\sum_{m=1}^{l}c_{m}\left\vert A_{j,\pi_{k}(m)}\right\vert ^{2}$ with
$c_{m}=\left.  \lambda_{\pi_{k}\left(  m\right)  }\overline{P}_{k}\right/
d_{j,\pi_{k}(m)}^{\gamma}$ for all $m=1,2,\ldots,l$, and $\overline{\theta
}_{k,l}$ is given by (\ref{A4_alltheta_defs}). The dominant term of each
$F_{H_{j,l}^{sum}}$ is proportional to $\overline{P}_{k}^{-l}$, and thus, the
dominant term of $1-F_{\Theta_{k,l}|\underline{\Theta}_{k}^{l-1}}$ is
proportional to $\overline{P}_{k}^{-l\left(  L_{k}-l\right)  }$.

For a fixed $R_{k}$, we lower bound $P_{o}^{(k)}$ by the outage probability
$P_{o,L_{k}\times1}$ of a $L_{k}\times1$ distributed MIMO channel in
(\ref{A3_Po_MIMO}). Generalizing the analyses in Appendix \ref{Comp_App2}, we
upper bound $P_{o}^{(k)}$ as (see (\ref{A4_Pout}) and (\ref{A4_I2_c}))%
\begin{equation}
P_{o}^{(k)}\leq\mathbb{E}\min_{l\in\mathcal{K}}(P_{o,l}^{(k)}(\underline
{\Theta}_{k}))=P_{UB}^{(k)}\label{A4_PkUB0}%
\end{equation}
where we use Lemma \ref{Lemma_Hypo} to write
\begin{equation}
P_{o,l}^{(k)}(\underline{\theta}_{k})\overset{\vartriangle}{=}\Pr\left(
G_{l}<\frac{R_{k}}{\theta_{k,l}}\right)  =\frac{\left(  2^{R_{k}/\theta_{k,l}%
}-1\right)  ^{l}}{\left(  l!\right)  \left(  \overline{P}_{k}\right)  ^{l}%
}\left(  \prod_{j=1}^{l}\frac{d_{d,\pi_{k}\left(  j\right)  }^{\gamma
}\overline{\theta}_{k,j}^{sum}}{\lambda_{\pi_{k}\left(  j\right)  }}\right)
+O\left(  \overline{P}_{k}^{-l-1}\right)  .\label{A4_Pol}%
\end{equation}
The probability $P_{UB}^{(k)}$ is given as (see (\ref{A4_Pout}) and
(\ref{A4_alltheta_defs}))%
\begin{equation}
P_{UB}^{(k)}=\int_{\theta_{k,1}=0}^{1}\int_{\theta_{k,2}=0}^{\overline{\theta
}_{k,2}^{sum}}\ldots\int_{\theta_{k,L_{k}-1}=0}^{\overline{\theta}_{k,L_{k}%
-1}^{sum}}P_{\underline{\Theta}_{k}}\left(  \underline{\theta}_{k}\right)
\min_{l\in\mathcal{K}}(P_{o,l}^{(k)}(\theta_{k,l}))d\theta_{k}%
.\label{A4_PUBMH}%
\end{equation}
For any $0<\theta_{k,l}^{\ast}<\overline{\theta}_{k,l}^{\ast}$, $1\leq
l<L_{k}$, the integral in (\ref{A4_PUBMH}) over the $\left(  L_{k}-1\right)
$-dimensional hyper-cube can be written as a sum of $2^{L_{k}-1}$ integrals,
each spanning $\left(  L_{k}-1\right)  $-dimensions, such that there are
$\tbinom{L_{k}-1}{j}$ integrals for which $j$ of the $\left(  L_{k}-1\right)
$ $\theta_{k,l}$ parameters range from $0$ to $\theta_{k,l}^{\ast}$,
$j=0,1,\ldots,L_{k}-1$ while the remaining range from $\theta_{k,l}^{\ast}$ to
$1$. Thus, we upper bound $P_{UB}^{(k)}$ in (\ref{A4_PUBMH}) by%
\begin{equation}
\int_{0}^{\theta_{k,1}^{\ast}}\int_{0}^{\overline{\theta}_{k,2}^{sum}}%
\ldots\int_{0}^{\overline{\theta}_{k,L_{k}-1}^{sum}}P_{\underline{\Theta}_{k}%
}\left(  \underline{\theta}_{k}\right)  P_{o,L_{k}}^{(k)}(\underline{\theta
}_{k})d\underline{\theta}_{k}+\int_{\theta_{k,1}^{\ast}}^{1}\int
_{0}^{\overline{\theta}_{k,2}^{sum}}\ldots\int_{0}^{\overline{\theta}%
_{k,L_{k}-1}^{sum}}P_{\underline{\Theta}_{k}}\left(  \underline{\theta}%
_{k}\right)  P_{o,1}^{\left(  k\right)  }(\underline{\theta}_{k}%
)d\underline{\theta}_{k}\label{A4Pub0}%
\end{equation}
where the dominant outage terms for $\theta_{k,1}\leq\theta_{k,1}^{\ast}$ and
$\theta_{k,1}>\theta_{k,1}^{\ast}$ are bounded by $P_{o,L_{k}}^{(k)}%
(\underline{\theta}_{k})$ and $P_{o,1}^{(k)}(\underline{\theta}_{k})$,
respectively. Furthermore, using the monotonic properties of $P_{o,l}^{\left(
k\right)  }$, the first term in (\ref{A4Pub0}) is bounded by $P_{o,L_{k}%
}^{(k)}(\underline{\theta}_{k}^{\ast})$ and the second term is bounded by
$\left(  1-F_{\Theta_{k,1}}\left(  \theta_{k,1}^{\ast}\right)  \right)
P_{o,1}^{(k)}(\underline{\theta}_{k}^{\ast})$. From (\ref{A4_Fthethak}) and
(\ref{A4_Pol}), using the fact that $P_{o,1}^{(k)}(\underline{\theta}%
_{k}^{\ast})$ has the smallest absolute exponents of $\overline{P}_{k}$,
namely $1$, and $\left(  1-F_{\Theta_{k,1}}\left(  \theta_{k,1}^{\ast}\right)
\right)  P_{o,1}^{(k)}(\underline{\theta}_{k}^{\ast})$ scales as $\overline
{P}_{k}^{-L_{k}}$, we bound $P_{UB}^{(k)}$ as
\begin{align}
P_{UB}^{(k)} &  \leq P_{o,L_{k}}^{(k)}(\underline{\theta}_{k}^{\ast})+\left(
1-F_{\Theta_{k,1}}\left(  \theta_{k,1}^{\ast}\right)  \right)  P_{o,1}%
^{(k)}(\underline{\theta}_{k}^{\ast})\\
&  \leq\frac{\left(  2^{R_{k}}-1\right)  ^{L_{k}}}{\left(  L_{k}!\right)
\left(  \overline{P}_{k}\right)  ^{L_{k}}}\left(  \prod_{j=1}^{L_{k}}%
\frac{d_{d,\pi_{k}\left(  j\right)  }^{\gamma}}{\lambda_{\pi_{k}\left(
j\right)  }}\right)  \cdot\left[  K_{c}+K_{d}\right]  +O\left(  \overline
{P}_{k}^{-L_{k}-1}\right)  \label{A4PubSum}%
\end{align}
where
\begin{equation}%
\begin{array}
[c]{ccc}%
K_{c}=\frac{\left(  2^{R_{k}/\theta_{k,L_{k}}^{\ast}}-1\right)  ^{L_{k}%
}\left(  \prod_{j=1}^{L_{k}}\left(  \overline{\theta}_{k,j}^{sum}\right)
^{\ast}\right)  }{\left(  2^{R_{k}}-1\right)  ^{L_{k}}} & \text{and} &
K_{d}=\frac{\left(  2^{R_{k}/\theta_{k,1}^{\ast}}-1\right)  ^{L_{k}}\left(
L_{k}!\right)  }{\left(  2^{R_{k}}-1\right)  ^{L_{k}}}\cdot\prod_{j=2}^{L_{k}%
}\frac{d_{\pi_{k}\left(  j\right)  ,\pi_{k}\left(  1\right)  }^{\gamma}%
}{\lambda_{\pi_{k}\left(  j\right)  }}.
\end{array}
\label{A4_KcKd}%
\end{equation}
Combining (\ref{A4PubSum}) with the lower bound in (\ref{A3_Po_MIMO}), we see
that the maximum achievable DDF diversity of a multi-hop TD-UC network is
$L_{k}$.

\bibliographystyle{IEEEtran}
\bibliography{MARC_refs}

\begin{thebibliography}{10}
\providecommand{\url}[1]{#1}
\def\UrlFont{\rmfamily}
\providecommand{\newblock}{\relax}
\providecommand{\bibinfo}[2]{#2}
\providecommand\BIBentrySTDinterwordspacing{\spaceskip=0pt\relax}
\providecommand\BIBentryALTinterwordstretchfactor{4}
\providecommand\BIBentryALTinterwordspacing{\spaceskip=\fontdimen2\font plus
\BIBentryALTinterwordstretchfactor\fontdimen3\font minus
  \fontdimen4\font\relax}
\providecommand\BIBforeignlanguage[2]{{%
\expandafter\ifx\csname l@#1\endcsname\relax
\typeout{** WARNING: IEEEtran.bst: No hyphenation pattern has been}%
\typeout{** loaded for the language `#1'. Using the pattern for}%
\typeout{** the default language instead.}%
\else
\language=\csname l@#1\endcsname
\fi
#2}}

\bibitem{cap_theorems:SKM04}
L.~Sankaranarayanan, G.~Kramer, and N.~B. Mandayam, ``Cooperation vs.
  hierarchy: An information-theoretic comparison,'' in \emph{Proc. {IEEE} Int.
  Symp. Inf. Theory}, Adelaide, Australia, Sept. 2005, pp. 411--415.

\bibitem{cap_theorems:KramMarYates01}
G.~Kramer, I.~Mari{\'c}, and R.~D. Yates, \emph{Cooperative
  Communications}.\hskip 1em plus 0.5em minus 0.4em\relax now Publishers, 2006,
  vol.~1, no. 3-4, pp. 271-425.

\bibitem{cap_theorems:Atheros_WP}
{Atheros Communications}, ``Power consumption and energy efficiency comparisons
  of wlan products,''
  www.atheros.com/pt/whitepapers/atheros\_power\_whitepaper.pdf.

\bibitem{cap_theorems:HienChandBala}
W.~Heinzelman, A.~Chandrakasan, and H.~Balakrishnan, ``Energy-efficient routing
  protocols for wireless microsensor networks,'' in \emph{Proc. 33rd Hawaii
  Intl. Conf. Systems and Sciences}, Maui, HA, Jan. 2000, pp. 1--10.

\bibitem{cap_theorems:AGS01}
K.~Azarian, H.~{El Gamal}, and P.~Schniter, ``On the achievable
  diversity-multiplexing tradeoff in half-duplex cooperative channels,''
  \emph{IEEE Trans. Inform. Theory}, vol.~51, no.~12, pp. 4152--4172, Dec.
  2005.

\bibitem{cap_theorems:JNL01}
J.~N. Laneman, ``Network coding gain of cooperative diversity,'' in \emph{Proc.
  IEEE Military Comm. Conf. (MILCOM)}, Monterey, CA, Nov 2004, pp. 3714--3722.

\bibitem{cap_theorems:LalithaSankar}
\BIBentryALTinterwordspacing
L.~Sankar, ``Relay cooperation in multiaccess networks,'' Ph.D. dissertation,
  Rutgers, The State University of New Jersey, New Brunswick, NJ, 2007.
  [Online]. Available: \url{http://www.winlab.rutgers.edu/$\sim$lalitha}
\BIBentrySTDinterwordspacing

\bibitem{cap_theorems:CTbook}
T.~M. Cover and J.~A. Thomas, \emph{Elements of Information Theory}.\hskip 1em
  plus 0.5em minus 0.4em\relax New York: Wiley, 1991.

\bibitem{cap_theorems:KGG_IT}
G.~Kramer, M.~Gastpar, and P.~Gupta, ``Cooperative strategies and capacity
  theorems for relay networks,'' \emph{IEEE Trans. Inform. Theory}, vol.~51,
  no.~9, pp. 3027--3063, Sept. 2005.

\bibitem{cap_theorems:ZT01}
L.~Zheng and D.~N.~C. Tse, ``Diversity and multiplexing: A fundamental tradeoff
  in multiple-antenna channels,'' \emph{IEEE Trans. Inform. Theory}, vol.~49,
  no.~5, pp. 1073--1096, May 2003.

\bibitem{cap_theorems:NBK01}
R.~U. Nabar, H.~B{\"o}lcskei, and F.~W. Kneub{\"u}hler, ``Fading relay
  channels: Performance limits and space-time signal design,'' \emph{{IEEE}
  JSAC}, vol.~22, no.~6, pp. 1099--1109, Aug. 2004.

\bibitem{cap_theorems:Cox01}
D.~R. Cox, \emph{Renewal Theory}.\hskip 1em plus 0.5em minus 0.4em\relax
  London: Methuen's Monographs on Applied Probability and Statistics, 1967.

\bibitem{cap_theorems:GKR01}
G.~Kramer, ``Models and theory for relay channels with receive constraints,''
  in \emph{42nd Annual Allerton Conf. on Commun., Control, and Computing},
  Monticello, IL, Sept. 2004.

\end{thebibliography}

\pagebreak%

\bigskip%

%

\bigskip%

\bigskip%

%

\end{document}